\documentclass[conference]{IEEEtran}

\IEEEoverridecommandlockouts
\usepackage[english]{babel}

\usepackage{cite}
\usepackage{amsmath,amssymb,amsfonts}
\usepackage{graphicx}
\usepackage{textcomp}
\usepackage{xcolor}
\def\BibTeX{{\rm B\kern-.05em{\sc i\kern-.025em b}\kern-.08em
    T\kern-.1667em\lower.7ex\hbox{E}\kern-.125emX}}

\usepackage{hyperref}
\usepackage[T1]{fontenc}
\usepackage{amsmath}
\usepackage{algorithm}
\usepackage[noend]{algpseudocode}
\usepackage{graphicx}
\usepackage[svgnames]{xcolor}
\usepackage{tabularx}
\usepackage{booktabs}
\usepackage{listings}
\usepackage{pdfpages}
\usepackage{subcaption}
\usepackage{caption}
\usepackage{xspace}
\usepackage{comment}
\usepackage{makecell}
\usepackage{multirow}

\usepackage{booktabs}
\usepackage{tikz}
\newcommand{\minibar}[2]{%
  \begin{tikzpicture}[baseline=-0.6ex]
    \fill[gray!25] (0,0) rectangle (1.8,0.14);
    \fill[blue!60] (0,0) rectangle ({1.8*#1/#2},0.14);
  \end{tikzpicture}%
}

\PassOptionsToPackage{hyphens}{url}
\usepackage{xurl} 

\newcommand{\AGILE}{AGILE US Government\xspace}
\newcommand{\SysName}{\pnamenospace\xspace}
\newcommand{\excise}[1]{}

\lstnewenvironment{PythonCode}[1][]{%
    \lstset{#1,float,language=python,%
      basicstyle={\fontfamily{lmtt}\selectfont\small},
      columns=fixed,                              % nice spacing
      xleftmargin=1cm,
      xrightmargin=1cm,
      numbers=none,frame=none,%
      keywordstyle={\bfseries\color{DarkRed}},%,          % keywords
      commentstyle={\itshape\color{DarkGreen}},%         % comments
      stringstyle=\color{black}%             % strings
    }%
  }%
  {}
\lstdefinelanguage{Julia}{morekeywords={abstract,break,case,catch,const,continue,do,else,elseif,%
      end,export,false,for,function,immutable,import,importall,if,in,%
      macro,module,otherwise,quote,return,switch,true,try,type,typealias,%
      using,while},%
   sensitive=true,%
   alsoother={\$},%
   morecomment=[l]\#,%
   morecomment=[n]{\#=}{=\#},%
   morestring=[s]{"}{"},%
   morestring=[m]{'}{'},%
}[keywords,comments,strings]%

\lstnewenvironment{JuliaCode}[1][]{%
  \lstset{#1,float,language=julia,%
    basicstyle={\fontfamily{lmtt}\selectfont\small},
    columns=fixed,                              % nice spacing
    xleftmargin=1cm,
    xrightmargin=1cm,
    numbers=none,frame=none,%
  }%
}%
{}

\newcommand{\aac}[1]{{\color{blue}(AAC: #1)}}

\newenvironment{tightcenter}{\par\noindent\begin{minipage}{\linewidth}\centering}{\end{minipage}}
\usepackage{enumitem}

\newcommand{\pname}{UpDown~}
\newcommand{\pnamenospace}{UpDown}
\newcommand{\lane}{lane~}
\newcommand{\lanenospace}{lane}
\newcommand{\Lane}{Lane~}

\newcommand{\lanes}{lanes~}
\newcommand{\lanesnospace}{lanes}

\newcommand{\ErdosRenyi}{Erd\H{o}s--R\'enyi\xspace}

\begin{document}

%\title{How Fast Can Graph Computations Go on Fine-grained Parallel Architectures}
\title{Fine-Grained Parallel Graph Algorithms Can Be Both Fast and Scalable}


% ORCIDs
% Alex: https://orcid.org/0000-0002-9955-2643
% Ivy: https://orcid.org/0009-0006-3548-5652
% Tianshuo: https://orcid.org/0000-0002-1275-2918
% Marzi: https://orcid.org/0009-0007-5124-8283
% Wenyi: https://orcid.org/0000-0002-0934-8770
% Jose: https://orcid.org/0000-0001-6875-1685
% Andronicus: https://orcid.org/0009-0005-1600-1459
% Jiya: https://orcid.org/0000-0002-4952-1486
% Jerry Xu: https://orcid.org/0009-0007-7269-9155
% Rajat: https://orcid.org/0009-0004-5452-3218
% Tony: https://orcid.org/0000-0002-7469-6130
% David: https://orcid.org/0000-0002-8107-6474
% Yanjing: https://orcid.org/0000-0003-0124-0463
% Hank: https://orcid.org/0000-0003-0816-8150
% Andrew: https://orcid.org/0000-0002-1204-206X

\author{\IEEEauthorblockN{1\textsuperscript{st} Yuqing Wang}
\IEEEauthorblockA{\textit{Department of Computer Science } \\
\textit{University of Chicago}\\
Chicago, USA \\
yqwang@uchicago.edu}
\and
\IEEEauthorblockN{2\textsuperscript{nd} Charles Colley}
\IEEEauthorblockA{\textit{Department of Computer Science} \\
\textit{University of Purdue}\\
West Lafayette, USA \\
ccolley@purdue.edu}
\and
\IEEEauthorblockN{3\textsuperscript{rd} Jiya Su}
\IEEEauthorblockA{\textit{Department of Computer Science } \\
\textit{University of Chicago}\\
Chicago, USA \\
jiya@uchicago.edu}
\and
\IEEEauthorblockN{4\textsuperscript{th} Brian Wheatman}
\IEEEauthorblockA{\textit{Department of Computer Science } \\
\textit{University of Chicago}\\
Chicago, USA \\
wheatman@uchicago.edu}
\and
\IEEEauthorblockN{5\textsuperscript{th} Alexander Fell}
\IEEEauthorblockA{\textit{Department of Computer Science } \\
\textit{University of Chicago}\\
Chicago, USA \\
alefel@uchicago.edu}
\and
\IEEEauthorblockN{5\textsuperscript{th} Ruiqi Xu}
\IEEEauthorblockA{\textit{Department of Computer Science } \\
\textit{University of Chicago}\\
Chicago, USA \\
ruiqix@uchicago.edu}
\and
\IEEEauthorblockN{5\textsuperscript{th} Andronicus Rajasukumar}
\IEEEauthorblockA{\textit{Department of Computer Science } \\
\textit{University of Chicago}\\
Chicago, USA \\
andronicus@uchicago.edu}
\and
\IEEEauthorblockN{6\textsuperscript{th} David F. Gleich}
\IEEEauthorblockA{\textit{Department of Computer Science} \\
\textit{University of Purdue}\\
West Lafayette, USA \\
dgleich@purdue.edu}
\and
\IEEEauthorblockN{7\textsuperscript{th} Andrew A. Chien}
\IEEEauthorblockA{\textit{Department of Computer Science } \\
\textit{University of Chicago}\\
\textit{Argonne National Laboratory}\\
Chicago, USA \\
aachien@uchicago.edu}
}
% \author{\IEEEauthorblockN{Anonymous Authors}}
% \IEEEpeerreviewmaketitle

%\tableofcontents
%\newpage

\maketitle

\begin{abstract}
Making graph algorithms fast and scalable on traditional architectures and GPUs often involves breaking them into large, coarse-grained computational blocks that access similar regions of memory. We show that graph algorithms, which have tremendous vertex- and edge-level parallelism, can be scaled as fine-grained programs on suitable architectures.  Specifically, we investigate graph algorithms on the UpDown architecture and show how this many-threaded, cache-less architecture enables simple, fine-grained parallel graph algorithms to be both fast and scalable. Single-node UpDown performance is 28x (geomean) over an H200 GPU, and that 
same program scales to 1000's of nodes.
We evaluate the system three ways: simulation, analytical projection models, and selected validation studies at scale (simulation of 16,384 UpDown nodes, using
256 supercomputer nodes). These results show the UpDown system can achieve 680K giga-traversed edges per second (GTEPS) for PageRank and 989K GTEPS BFS, exceeding
prior results significantly.

\end{abstract}


\section{Introduction}
\label{sec:introduction}

High-performance graph computing is a key challenge for computer systems because most graph algorithms for the very large graphs have linear or mild superlinear complexity. This situation tends to indicate limited opportunity to reuse data transported from memory to processor caches, or that doing so involves complex access strategies and algorithmic formulations. Moreover, degree irregularity and effectively random data access patterns render prefetching less effective. Beyond these challenges with memory and data access, graph algorithms are often conceived of in a ``think like a vertex'' fashion~\cite{Malewicz-2010-Pregel,Gonzalez-2012-powergraph}. This provides fine-scale, vertex-level parallelism that is often challenging to scale in distributed settings. The shared memory, scale-up scenario is slightly easier~\cite{Shun-2013-ligra,McSherry-2015-cost}  with known results about optimal algorithms and software frameworks to match~\cite{dhulipala2021theoretically}. However, this leaves a significant performance gap between scale-out graph algorithms on distributed systems and scale-up graph algorithms on shared-memory systems. The improved memory bandwidth of GPUs creates yet another gap between the performance of systems where the graph fits into GPU memory. 

This scenario results in a wide diversity of research into high-performance graph computations spanning algorithms, data layouts, frameworks, graph databases, and system architectures. Here, we investigate a variety of graph algorithms on a newly proposed computer architecture called UpDown with the aim of looking at the question: \emph{can fine-grained graph algorithms be fast (efficient) and scalable (to supercomputer scale)?}
%it might be possible for graph algorithms to go on a distributed system designed for sparse irregular data access}?

%We find that this proposed multinode system architecture, evaluated via cycle accurate processor simulation, demonstrates performance that exceeds current systems where the graph fits into GPU memory entirely. 
%Although not a direct focus of this paper, we note that the system enables scalable computations with ``think like a vertex''-like paradigms. 

To answer the question, we first describe the design of a novel parallel architecture, the \SysName system \cite{ChienUpdownTPDS26,Single-Chip-UpDown26,UnlimitedMP24}.  Research efforts under the \AGILE program~\cite{agileprogram} created a detailed hardware design, programming and simulation tools, enabling study of \SysName on graph applications at scale.\footnote{With regards to anonymity, as part of the AGILE program, UpDown programming tools, simulators, and materials were distributed to the Test and Evaluation teams as well as all of the performers to enable them to test algorithms on UpDown. Any of these entities had access to materials sufficient to produce the results in this paper. }  %novel architectures for graphs. 
%IARPA AGILE program and driven to detailed design and studied with extensive simulation \cite{AGILE}.  
We first study seven diverse graph algorithms, demonstrating the breadth of \pnamenospace's performance benefits.  Next,
we conduct a deep study of two fundamental kernels, PageRank~\cite{page1999-pagerank} (PR) and breadth-first search (BFS).  Breadth-first search is a purely discrete algorithm, whereas PageRank models a simple numerical iteration over the edges of a graph. Both exhibit low data reuse and linear computational complexity, making it difficult to achieve high performance.  When used on graphs such as RMAT, the extremely skewed degree distributions create a large work imbalance and effectively random data access patterns. 

%\aac{add discussion of small eval}

To evaluate whether fine-grained parallel algorithms can be fast (efficient), we conduct a broad evaluation of seven algorithms and two kernels, comparing their performance against both CPU and GPU implementations.  These results evaluate the premise that fine-grain parallelism can be fast.  

Next, we shift to assessing scalability. We evaluate medium-scale UpDown systems with over 500,000 independent computing units. These simulations are detailed, with instruction-level timing accuracy, network latency, and memory bandwidth limits.  To assess achievable performance for larger systems, we combine simulation data with analytical models of the graphs and algorithms to project performance for full-scale \pname system designs (33 million-fold parallelism, 16K nodes).  Using a supercomputer to simulate all 33 million compute units, we validated two of the projected runtimes on an 8.6 billion-vertex graph with 265 billion undirected edges. The supercomputer results matched well with per-iteration runtimes for PageRank (0.58ms (projected)) vs. 0.65ms and BFS (0.91ms (projected)) vs. 0.84ms. 

Finally, we compare \pname system performance to the fastest published results on today's supercomputers, comparing scalability and absolute performance.  Specific paper contributions include:

%% I removed the "measured" because that to me implies there's a hardware artifact

% \david{Need info for BFS!}.  %These studies project both absolute performance achieved, and ISO-power normalized performance.
% \textbf{\textcolor{red}{XXXX-Needs update above to be updated based on the final figure-XXXX}}

%one of the \AGILE~\cite{agileprogram} program's goals 

%Finally, programming scalable graph applications is difficult, and a key goal is to combine programmability with efficient performance.  Thus, %To show the potential of this novel architecture, to showcase \SysName programmability, we implemented five different algorithmic variants of PageRank and BFS, demonstrating both efficient performance and excellent scalability for each.
%In graph computations, giga/billions of traversed edges per second (GTEPs) is a widely used metric. 

%These were programmed with modest effort and achieve significant performance increases.
%beyond that provided by the improved computing hardware.

%\brian{this paragraph feels a bit off, as said above, scalability isn't hard, efficiency is hard, you can implement PR and BFS pretty easily with off the shelf MapReduce and get scalability, but bad absolute performance.  Also, this seems to indicate that the performance of variants shows something about programmability, its not their performance, its how difficult they were to program. }

%goal: to frame the ambition of application scientists, algorithmists, and software as well as computer architects

%evaluate pr and bfs on it

%show performance and benefits of programmability

%\todo{Finish contributions}

\begin{itemize}[leftmargin=*]
\item A broad study of graph algorithms shows expression and exploitation of fine-grained parallelism enables a 1-node UpDown to outperform an H200 GPU %at generally more than 1000x faster than a CPU, and 
by 28x (geomean). % faster than an H200 GPU.
%    \item novel programming approach that exploits natural expression of graph computing algorithms -- directly for programmabilty; and direct exploitation of this fine-grained parallelism in hardware
    % \item Using a direct {\it natural} expression of vertex and edge parallelism, evaluation of BFS and PageRank (fine-grained), using a fine-grained architecture designed for fine-grained event-driven threads, shows 2000 to 5000-fold speedup over a single x86 core and 1 million GTEPs performance for projections to larger scale systems and graphs. \brian{what is the comparison which is getting 2000-5000x speedup, is it per lane, equal area, equal power?}
    \item 
    % Extremely large amounts of parallelism can be expressed using a direct {\it natural} expression of vertex and edge for PageRank and BFS. 
    %Simulations show the \pname fine-grained architecture %, designed with event-driven threads, %exploits this parallelism efficiently.
    Simulations show scale-out performance.  A 256-node \pname achieves speedups of up to 304x (PageRank) and 377x (BFS).
    %    (self-relative, 256 nodes over 1 node). For 256 nodes,
    Reaching absolute performance of 12,850 GTEPS (PageRank) and 18,231 GTEPS (BFS).
    
    \item Projections indicate a full-sized \pname system (16,384 nodes) would reach 684K GTEPS on 
    PageRank (Graph500 RMAT Scale 40, $|V| = 2^{40}$). Exploiting \pnamenospace's programmability, we studied 
    %enables the use of a
%    fine-grained architecture and programming model enables easy modification of 
a better algorithm, data-driven PR, improving performance 14-fold. 
%increase efficiency.  For example, employing a work-reducing variant of PageRank can 
%improve performance by up to 14x. %4-fold. 
For BFS, \pname reaches 990K GTEPS on 16,384 nodes.

\item We validate the projections, with Graph500 Scale 33 RMAT simulations of a 
16,384-node \pnamenospace.  These simulations run on 256 Perlmutter supercomputer nodes, employing over 110 TB of memory,
%These studies on Graph500 Scale 33 RMAT %Graphs ($|V| = 2^{33} = 8.6$ billion vertices) 
validate the projection models, with 
compute times in close agreement, increasing confidence.

    %\item Network modeling shows that the \pname design supports up to 8K nodes, but an increase in link speed is required for scaling to 16K nodes.
    
    %and can compute  \brian{can this be compared to the current graph 500 numbers?  there is a question of if that comparison is done here or in the last bullet}
    %\brian{scaleX graph needs to be defined}
    %\item \ivy{@charlie add performance claim for projection }
    % \item Evaluation of algorithmic variants of BFS and PageRank (fine-grained), using a fine-grained architecture demonstrating the flexible programmability, shows as much as a 2-fold performance increase.  \brian{same as above, how does performance show programmability} 
    %\ivy{add claim for data driven}
    %\item The fine-grained architecture and programming model enables use of simple algorithms and easy elaboration/tuning for higher performance.  For example, a more work efficient variant of PageRank that  \todo{doubles, 10x?} the absolute performance and push, pull, and self load-balancing variants of BFS %\todo{?} for skewed graphs.  
    %\item \jiya{parallelism, utilization, network latency in section \ref{sec:performance-comes-from}?}
    \item Comparing absolute performance to existing supercomputers shows \pname can increase performance by 10-100x.
    %\jiya{Lack 1-node performance comparison with H200 GPU.}
    
%    show UpDown is 345x faster on PageRank, improving to 4,830x if better algorithms are used.  \pnamenospace's BFS performance is 5x long-time Graph500 leader Fugaku (10x ISO-power)  %and without aggressive preprocessing, 
%    and 2.3x current Graph500 leader NVIDIA EOS GPU-based system (and we expect 4.6x with software optimization). 
%    \jiya{someone might ask why PR speedup is so high. Lack 1-node performance comparison with GPU.}

    % \aac{For David: Need to figure out what to say about new eos-dfw result} %\brian{ needs cites}.
%    5x than Fugaku on BFS (w/o preprocessing), 25x EOS nvidia

    % increases as high as 2000-fold, and this performance is accessible on much smaller graphs. %iso-power efficiency
\end{itemize}

%\david{TODO: Make sure this is consistent...}
The paper is organized as follows: 
%in Section~\ref{sec:background}, we present background, discussing parallel architectures, parallel PageRank and BFS approaches, along with other approaches to large graph computation. 
in Section~\ref{sec:approach}, we describe the approach and the \SysName architecture and programming model. %and its support for fine-grained parallelism. 
Section \ref{sec:small-eval} studies fast (efficient) graph algorithms on 1 node. Then scaling to 256 \pname nodes.  Next, in Section~\ref{sec:evaluation}, we
evaluate 
%describe the simulation methodology to evaluate 
the \SysName architecture,
%We also summarize the implementations of PageRank and BFS, as well as a work-reducing variant. The results discussed 
%including (i) simulated 256-node performance and scalability,
%characterized with detailed simulation, 
project performance for a full 16,384-node system, and validate it with a supercomputer-based simulation. 
%We also study a more detailed network evaluation to augment the other simulation data. 
% \aac{For David: Update this} 
%that confirms the assumptions of simulation and projection. 
Section \ref{sec:performance-comes-from} explains where \pname performance comes from.
Section~\ref{sec:other} discusses \pname performance in context. 
%to other scalable systems.
%, presenting the absolute performance increase \pname shows that codesign can achieve. 
Finally, Section~\ref{sec:summary} summarizes and proposes future directions. 
%with a discussion of related systems and list future research directions in Section~\ref{sec:summary}.

%\newpage

\section{Approach: How UpDown Enables Scalable Graph Computations}
\label{sec:approach}

% \aac{
%  (Andrew outline) revise section 3 to point out the weaknesses of other parallel graph architectures (keep at same length, 
%    cut down architecture description (can we put the full desc in an appendix? point to existing publications for 
%    description)
%    a. accelerators, → poor scaling / no scaling story
%    b. cpu's and gpu's poor datapath utilization → caches require hero programming effort to utillize for performance. 
% }

Graph processing is difficult for %\brian{modern or traditional? if feels like a word is missing here} 
computer architectures because it has low data reuse, eliminating the benefits of caches and deep memory hierarchies.  Worse, references are often sparse, so cache block transfers (typically 64 bytes) can waste 7/8 (87.5\%) of the data movement.  

\begin{figure}
    \centering
    \includegraphics[width=0.95\linewidth]{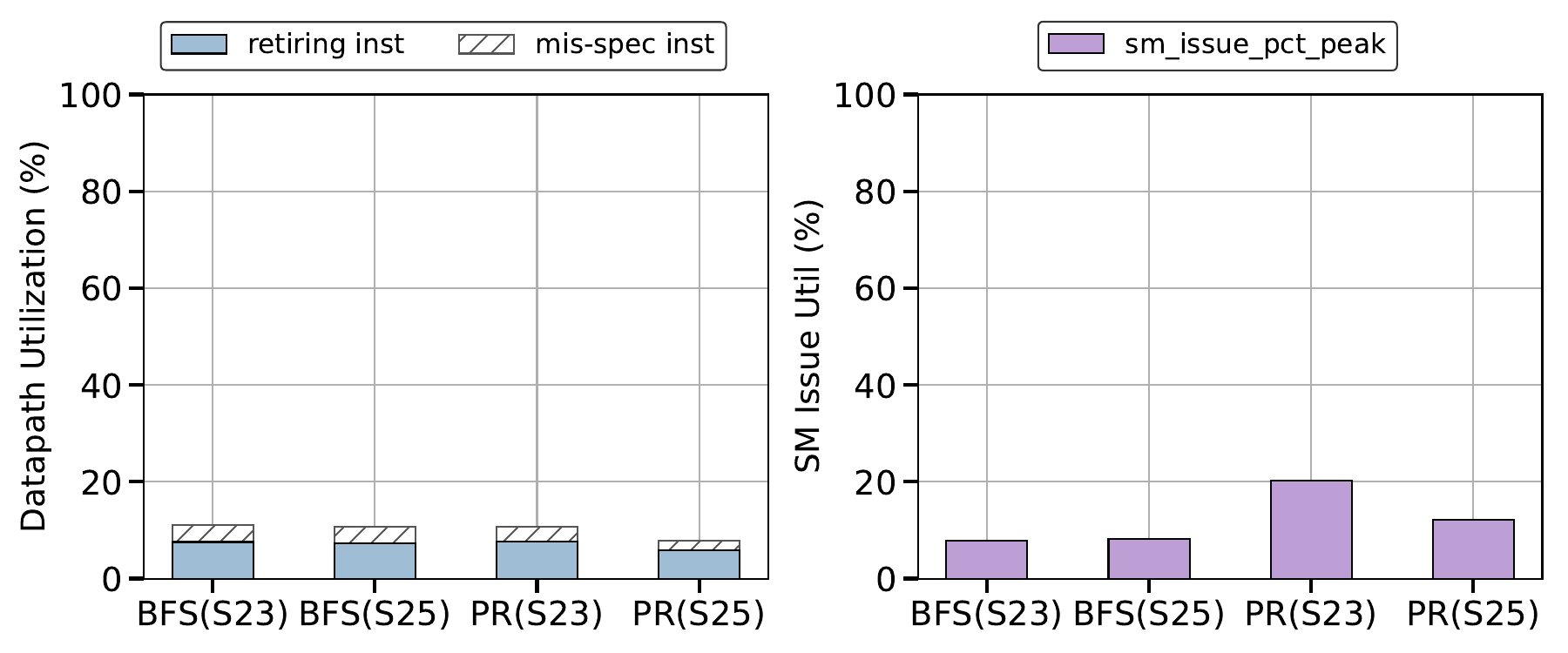}
%    \caption{Datapath Utilization for CPU and GPU on graph algorithms, from \cite{SingleChip26}}

\caption{Datapath Utilization for BFS and PageRank on %irregular, highly-skewed 
RMAT graphs. a) Core utilization on a 20-core Skylake CPU \cite{vtune, beamer_gap_2017} are $<10\%$, b) SM Issue slot utilization on a NVIDIA-A40 GPU are $< 20\%$ \cite{gunrock, ncu}; from \cite{Single-Chip-UpDown26}.}
    \label{fig:datapath-utils-cpu-gpu}
\end{figure}

Together, these effects lead to low datapath utilization in both CPUs and GPUs as shown in Figure \ref{fig:datapath-utils-cpu-gpu}.  The poor datapath and cache performance produce both low performance and poor power efficiency.

\subsection{Accelerator and Traditional Supercomputer Approaches}

These properties have given rise to many innovative proposals for specialized graph accelerators as surveyed \cite{gui_survey_2019}.  Many of these architectures are single-chip, often not scaling to large memory, and none scale out to supercomputer performance.  

%\aac{what about the Wuhan apsp machine?}.

Supercomputer systems with tightly integrated networks have achieved higher absolute performance on benchmarks such as breadth-first search (BFS), including record-setting Graph500 results~\cite{Murphy-2010-graph500}, but often at substantial programming cost.
First, for efficient single-node performance, significant programmer effort is required to optimize data layouts and traversals \cite{wheatman2021parallel, wheatman2024batch, wheatman2024byo, wheatman2021streaming, pandey2021terrace}.  Second, extensive  MPI-based distributed-memory optimization is required ~\cite{Fugaku-graph500,Gan2022,Gan2025}.  And ultimately, because they are scaling on CPUs and GPUs, they have even lower datapath utilization than that shown in Figure \ref{fig:datapath-utils-cpu-gpu}.

%\ivy{@Andrew these 3 paragraphs}

%and the many twisted ways programmers optimize for it (blocking, optimized traversals, edge-ghost-zones, etc.).  

%\aac{accelerators/limitations and see related work; scaling problems}

%\aac{scaling graphs on traditional computers and see related work; cpu/gpu building blocks}

%Further, CPUs and GPUs require large chunks of computation for efficiency so the schedulers or runtime must aggregate vertex and edge level logical operations into large chunks of work to be scheduled, which can substantially limit the amount of parallelism and the maximum possible speedup.

\subsection{UpDown Architecture Approach}
\label{sec:updown-architecture-approach}

The UpDown architecture is part of IARPA's AGILE program \cite{agileprogram}.  UpDown's system, architecture, performance, and implementation feasibility are well-described elsewhere \cite{ChienUpdownTPDS26,Single-Chip-UpDown26,UnlimitedMP24,rajasukumar2024updown,Andronicus-thesis}. Here, we summarize the key features.

%\brian{as written this is just wrong, most common in-memory frameworks exploit both vertex and edge parallelism, in fact I did a whole study in what happens if you don't exploit edge parallelism here https://www.vldb.org/pvldb/vol17/p2307-wheatman.pdf and found that on just 64 cores you get up to a 2x slowdown in practice. }

%we have built a simple programming system that expresses edge and vertex level parallelism and maps it onto a fine-grained architecture.

%examples of edge parallel, and vertex parallel operations

%examples of scalable abstractions (sets, attributes) that enable the expression of full algorithms

% \begin{figure}
%     \centering
%     \includegraphics[width=1\columnwidth]{figures/updown-arch6.png}
%     \caption{UpDown Lane Architecture}
%     \label{fig:updown-lane}
% \end{figure}

\begin{figure}%[t]
    \centering
    \includegraphics[width=1.1\columnwidth]{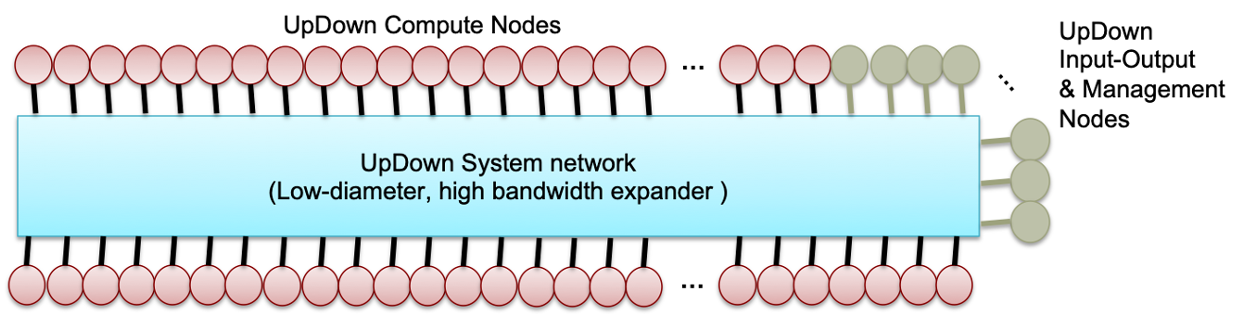} ~~~

    \caption{The \pname System has 16K nodes connected by a high-performance system network (latency: 500ns, node BW: 4.4TB/s, system bisection: 32 PB/s.  Each node includes a CPU, 2048 UpDown lanes, and 8 HBM DRAM stacks.}
    \label{fig:updown-system-node}
\end{figure}

\subsubsection{Fine-grained Parallelism and Low Data Reuse}
\label{sec:system_description}

% \begin{itemize}[leftmargin=0.125in]
% \item \emph{Fine-grained parallelism.}

The \pname system uses % compute building block is an 
a custom \lane that incorporates
mechanisms for efficient event-driven threads, split-transaction memory operations, and efficient short threads.  They also include hardware multithreading and efficient message send instructions.  
These mechanisms enable software to enjoy the execution costs in Table \ref{tab:lane-execution-costs}.  Each lane includes a 2KB I-cache and 64KB scratchpad (no data cache).
%\item 

\subsubsection{Massive MIMD parallelism and memory bandwidth}  An \pname node has 2,048 lanes, and a system has 16,384 nodes.
Each \lane has full MIMD flexibility at 2GHz, so the 33 million \lanes have a throughput of over 66 million billion instructions per second, or peta-instructions per second.  
Lanes are organized into accelerators, each with 64 lanes, and a cluster of four accelerators is associated with an HBM3e stack (1.2 terabytes/second). With 8 clusters per node, the system has a total memory bandwidth of 157 petabytes per second.
% (see Figure \ref{fig:updown-system-node}).

%\end{itemize}

\subsubsection{Flexible Data Layout and Use} %Programs leverage the 
The hardware-supported global address space enables programs to access DRAM across the entire machine directly via messages \cite{UpDownTrans24}.  Each load/store operation on 1-8 64-bit words costs 2 cycles overhead (1 to issue, 1 for response). %The latency from request to response is 300-1000's of cycles.  Within a node, all 2,048 \lanes can access all node DRAM with less than 150ns latency.  
System memory capacity is 8.4PB with $>$150 PB/s bandwidth, globally addressable.  Latency is 150 ns (300 cycles) within a node and 1,250 ns (2,500 cycles) internode, round-trip.  All 33 million \lanes can access the entire 8.4PB DRAM.
%with a round trip 1,250ns %\jiya{(1100ns in Fastsim2)} 
%latency.  
This is achieved with a diameter-3, low-latency global system network based on an enhanced version of Polarfly called PolarStar \cite{lakhotia2022polarfly,lakhotia2024polarstar}.  Each node has 4.4 TB/s injection bandwidth and 4.4 TB/s incoming bandwidth, and the system has 32 PB/s of bisection bandwidth.  \pname maximum system power is estimated  at 9.4 MW~\cite{ChienUpdownTPDS26,Single-Chip-UpDown26}.

System and \lane performance characteristics are based on the detailed \pname design that includes full architecture simulation and RTL (synthesis and timing/area/power estimation) ~\cite{ChienUpdownTPDS26,Single-Chip-UpDown26, UnlimitedMP24}.

% {rajasukumar2024updown,KVMSR24,,UpDownTrans24}.  

%The low-cost message and remote memory access costs (Table \ref{tab:lane-execution-costs}),

%This reflects higher global bandwidth / node enabling programs to use data flexibly, as is common in graph computations. 

\subsection{UpDown Programming Approach}

UpDown enables flexible global memory programming.  Programming is based on the UDWeave event-driven language, and parallelism is organized using a fine-grained map-reduce structure over a global address space\cite{Fell2025,KVMSR24}.

Applications for UpDown naturally express vertex- and edge-level parallelism, creating massive fine-grained parallelism.
This approach utilizes massive memory and system network bandwidth to achieve efficient, scalable performance.
%largely ignore data reuse and yet still 
This is possible because the overall system has a radically different communication-compute balance as shown in Table \ref{tab:comm-compute-balance}.
With this approach, the available fine-grained vertex and edge parallelism exceeds billion-fold as shown in Figure \ref{fig:thread-invocations}, and 
reaches trillion-fold for the largest graphs studied.

\begin{table}[t]
\begin{center}%\brian{load store being marked at cost of 2 feels off since its a throughput thing not a latency thing, maybe this could be expressed more clearly}}
\resizebox{.8\columnwidth}{!}{%
\begin{tabular}{@{}lrr@{}} \hline
Operation  & Instructions & Cost (cycles)  \\ \hline
Thread Create & 0 & 0  \\ 
Thread Yield & 1 & 1 \\
Thread Deallocate & 1 & 1 \\ 
Send Message & 1 & 1-2 \\ 
Issue Load/Store to DRAM & 2 & 2 \\ \hline
%Load/store (scratchpad) & 1 & 1 \\ \hline
\end{tabular}%
}
\caption{\Lane Execution Costs (2GHz clock)} 
\label{tab:lane-execution-costs}
\end{center}
%\vspace{-\baselineskip}
\end{table}

%The \pname system is fully described in \cite{anonUDSystem}, along with a broad evaluation across a range of computational kernels.
% The \pname ISA is available for the purposes of anonymous review from  \url{https://drive.google.com/file/d/14XovYu4ZiRQx0JwQ8u5PgU2wW5nJshFK/view?usp=sharing}
%\aac{Explain KVMSR+UDWeave, cite the paper}
%{\bf Fine-grained thread support in each \lane allows full exploitation of edge and vertex parallelism, using independent threads. Table \ref{tab:lane-execution-costs} shows the low costs for thread and messaging operations.}  

% \aac{add 1 picture from kvmsr+udweave paper?}

\begin{figure}[h]
    \centering
%    \aac{Updown software picture}
   \includegraphics[width=0.9\linewidth]{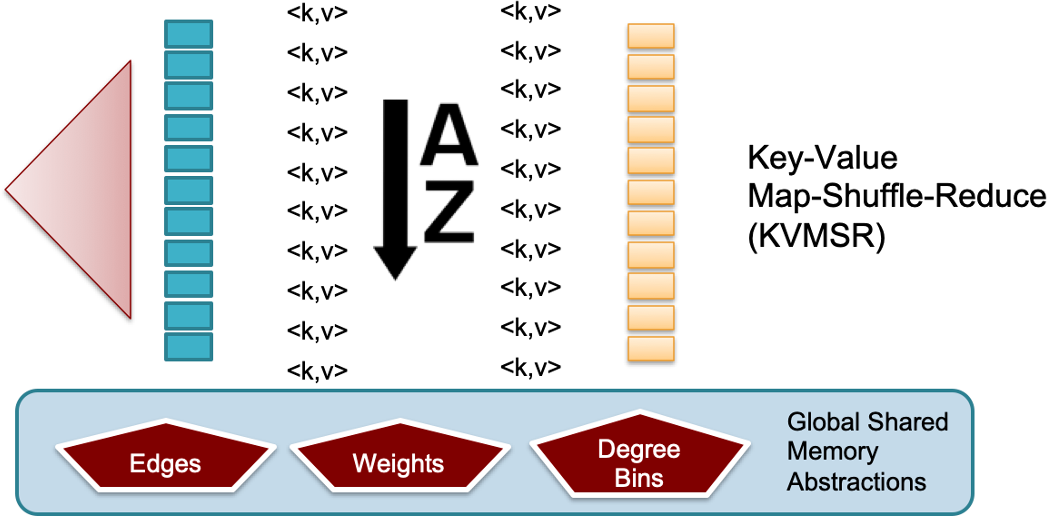}
    \caption{UpDown Software: KVMSR organizes parallelism over mutable data abstractions.  UDWeave is used to express fine-grained, event-driven tasks.}
    \label{fig:updown-software}
    \label{fig:kvmsr}
\end{figure}

The \pname programming system (KVMSR) employs the map-reduce paradigm, with very fine-grained map/reduce tasks, over a global shared memory \cite{Fell2025}.  Fine-grained, event-driven tasks are expressed in UDWeave.  These, combined with the global namespace, allow separate optimization of compute and data placement and exploit the \pname hardware for efficient execution (10-instruction tasks are efficient).  Load balance is achieved through hashing and graph restructuring (vertex splitting) to manage graph skew, enabling both efficiency and high absolute performance.

\begin{table}[t]
    \centering
    \resizebox{\columnwidth}{!}{ 
%    \aac{fix per-node, add per-socket column}
    % \begin{tabular}{p{0.25in}|r|r|p{0.42in}|p{0.43in}|p{0.49in}|p{0.43in}} 
     \begin{tabular}{@{}lrrrrrr@{}} 
    \toprule
     & Nodes & Sockets & \begin{tabular}{@{}c@{}} Node \\ Injection  \end{tabular} & \begin{tabular}{@{}c@{}} Per-Node \\ Bisection  \end{tabular} &\begin{tabular}{@{}c@{}} Per-Socket \\ Bisection  \end{tabular} & \begin{tabular}{@{}c@{}} System \\ Bisection \end{tabular} \\ 
     \midrule
    \pname & 16,384 & 16,384 & 4.4 TBps %& 64 PBps 
    & 2 TBps & 2 TBps & 32 PBps \\ 
    % \hline
    Aurora & 10,624 & 84,992 & 0.2 TBps %& 2.12 PBps 
        & 0.07 TBps & 0.014 TBps & 0.7 PBps \\ 
    \midrule
    Ratio (F/A) & 1.6 & 0.19 & 20 %& 30.2 
    & 29 & 148 & 46.4  \\ 
    \bottomrule 
%    Ratio (Aur/UD) & 0.64 & 5.2 & 5.85 & 0.05 & 0.033 & 0.035 & 0.022\\ \hline
    \end{tabular}
    }
    \caption{Comparing System Balance of \pname and Aurora Supercomputer  \cite{Aurora}.  \pname has 46x higher network bisection, and Aurora is 5.5x higher system power.}
    \label{tab:updown-aurora-comparison}
    \label{tab:comm-compute-balance}
\end{table}

% \fi

%Key performance attributes of the machine include a high degree of multithreading in each lane with 1 cycle thread creation and termination, massive fine-grained MIMD lanes each with 64KB scratchpad memory for fast access, and a globally addressed memory with low latency -- 70ns within a stack and 150ns to remote stacks. 

%\paragraph{Overall Approach}
%\aac{move this paragraph up? explain software level view, relegate hardware view to "completeness"}
%The \pname system %eliminates the programming effort radically reduces the programming effort to tune the graph computation to match the machine.  This is because system's hardware properties enable efficient exploitation of fine-grained parallelism directly (computations as small as 10 instructions), and simple direct access to a shared global memory (with ample global network bandwidth).  

%\textbf{The challenge of extreme performance is to show that this parallelism can be exploited, at acceptable programming effort.}  We exploit this directly

% \aac{Where does this go?}
% We use system simulation and performance modeling to evaluate achievable PageRank and Breadth First Search performance on highly-skewed graphs.  % on architectures such as \pname that are designed for extremely irregular and fine-grained parallelism.  
% These studies aim to show the performance potential of codesigned fine-grained architectures for graph computations.

\section{Diverse Algorithm Evaluation}
\label{sec:small-eval}

% \aac{(Andrew draft)  add a new section between sections 3 and 4, called "small scale evaluation", which includes a set of results
% from the "application templates".  k-truss, k-cores, tc, ?.   point out the comparative performance, point out the
% value for interactive, analyst usage modes. Maybe 1.5 pages}

% We implemented and studied a wide range of graph algorithms: triangle count~\cite{HPEC-ktruss,GBBS}, k-truss~\cite{HPEC-ktruss}, k-cores~\cite{kcore-paper, GBBS}, connected components~\cite{HPEC-lb}, strongly connected components~\cite{SCC-coloring}, Louvain~\cite{louvain, parallel_louvain}, breadth-first search~\cite{HPEC-lb, push-pullBFS}, and PageRank~\cite{HPEC-lb, whang2015scalable}. Across all of these algorithms, UpDown robustly outperforms both GPU (H200) and multi-core CPU systems.  UpDown also scales up to large-parallel configurations of 100's to 1000's of nodes.  Finally, UpDown supports real-time analytics within short deadlines.  

We implemented and studied a wide range of graph algorithms: triangle count, k-truss, k-cores, connected components, strongly connected components, Louvain, breadth-first search, and PageRank~\cite{HPEC-ktruss,GBBS, kcore-paper,HPEC-lb, SCC-coloring, louvain, parallel_louvain, push-pullBFS, whang2015scalable}. Across all of these algorithms, UpDown robustly outperforms both GPU (H200) and multi-core CPU systems.  
Finally, UpDown supports real-time analytics within short deadlines.  

%\aac{drop? UpDown also scales up to large-parallel configurations of 100's to 1000's of nodes.} 

\subsection{Summary of Algorithm Implementations}
For each computation, we briefly explain the implementation on UpDown. 

\textbf{Triangle Counting (TC).}
% We parallelize over the full set of intersections among vertex lists for each edge. 
% We use the standard degree-based edge orientation. 
% (We use split vertices to handle high degree vertices ... \david{Do we actually do this?}
We parallelize over edge-based vertex list intersections. As is standard, vertices are numbered in descending order of degree, and the standard degree-based edge orientation is applied to reduce comparisons. 

\textbf{K-Truss.} 
% The $k$-Truss implementation is explained in more detail in \cite{HPEC-ktruss}. 
K-Truss extends triangle counting by maintaining the support (the number of triangles) for each edge using atomic operations on values stored in the scratchpad and DRAM. Edges with support less than $k-2$ are iteratively removed until convergence. A previous $k$-Truss implementation on this architecture is detailed in \cite{HPEC-ktruss}.

\textbf{K-Core.}
The k-core algorithm identifies subgraphs where each vertex has degree at least $k$~\cite{kcore-paper}. The implementation for UpDown successively removes vertices of degree smaller than $k$ via load-balanced graph computations using~\cite{HPEC-lb}. This involves a sequential iteration through all values of $k$. Vertex degree is used as a workload estimate, and vertices (along with their incident edges) with degree less than $k$ are iteratively removed until convergence.

\textbf{Connected Components (CC).} 
Connected components are also implemented using the graph load-balancing framework \cite{HPEC-lb}. Each vertex is initially assigned a unique label equal to its vertex ID. In each iteration, vertices update their labels to the minimum label among their neighbors. Through iterative label propagation, all vertices within the same connected component converge to a common minimum label.

\textbf{Strongly Connected Components (SCC).}
SCC is computed on a directed graph using an iterative forward--backward traversal approach \cite{SCC-coloring}. We first apply a label propagation step similar to CC to partition the graph into candidate subsets. For each subset with label $c$, we select the representative root vertex $v_c$ (with $\mathrm{vid}=c$) and mark it as visited. We then perform one backward traversal (BFS on the reverse graph) from all visited roots to identify vertices that can reach their corresponding roots and share the same label. The resulting vertices form multiple SCCs, which are then removed. This process repeats until all vertices are assigned to SCCs. Both label propagation and BFS are implemented using the graph load-balancing framework \cite{HPEC-lb}.

\textbf{Louvain.} 
The Louvain algorithm consists of two phases: local move and graph aggregation. We found the local-move phase accounts for more than 95\% of the total runtime and focus on this phase. We implement a synchronized local-move variant using the graph load-balancing framework \cite{HPEC-lb}. Given an undirected CSR graph, the vertex degree is used as a workload estimate. For each vertex, we compute the modularity gain of moving it to neighboring communities and select the one that yields the highest gain.

We discuss \textbf{BFS} and \textbf{PageRank} in subsequent sections, along with deeper investigation and analytical performance modeling to synthesize performance at full system scale.

All the implementations derive their parallelism from these fine-grained structures naturally exposed in the graph data, along with a small set of parallel primitives: prefix sum, broadcast, reduction, and load balancing. For instance, k-core uses a prefix-sum and broadcast library, in addition to load-balancing tools.  Across most algorithms, the graph load-balancing framework distributes work (vertices and edges) based on workload estimation, allowing each thread to process multiple vertices or partial edge lists of high-degree vertices to improve load balance and resource utilization. Algorithms are implemented in the UDWeave programming language using a MapReduce-like library specialized to the UpDown architecture~\cite{Fell2025}.

\subsection{Algorithm Implementation Evaluations}
\label{subsubsec:modeling}
%simulation methodology (description of cycle accurate simulation for nodes, memories, using gem5)

%\aac{rewrite to reflect fastsim2 \\don't need system description (covered above)}
% \label{sec:simulator}

\pname performance on these algorithms is simulated with a cycle-accurate instruction-level simulator (Fastsim2) for each \pname \lanenospace.  It incorporates latency and rate models for all memory and internode communication systems. The simulator runs fast enough to enable 256-node (524K \lanesnospace) studies on a single multicore server. %graph computations. 
Fastsim2 was validated against a detailed  GEM5~\cite{gem5_simulator} simulation model
%The GEM5 model is a cycle-accurate hardware simulator.
that includes a DRAMsim3 model for the HBM stacks \cite{li_dramsim3_2020}.  
We also evaluate using a more scalable Fastsim3 simulator, as described in Section~\ref {sec:validation}.

\subsection{Graphs}
We use undirected graphs from the SNAP collection~\cite{snapnets}, Erd\H{o}s R\'{e}nyi (ER), Graph500 RMAT~\cite{Murphy-2010-graph500}, and ForestFire graphs~\cite{leskovec2007graph}. Both RMAT and ForestFire graphs exhibit substantial skew in their degree distributions, a key challenge for load balancing on parallel machines.
%and the skew in RMAT is particularly extreme. 
ForestFire graphs additionally model local structure in the graphs~\cite{Leskovec-2009-community-structure}. ER graphs are the best expression of pure randomness with \emph{no structure}.  With a uniform degree distribution, they are easier to load balance.
%and so they lack the challenge of the highly skewed degree distribution. 
If the generators produce directed edges, we simply remove the directions to produce an undirected graph. We compute ER graphs using $p = 35/n$ (average degree of 35), RMAT graphs use the Graph500 specification ($a = 0.57, b=c = .19$, and $d_{average} = 16$), and the forest fire graphs use $p_{burn} = .4$ (both in and out edges).  We drop the direction on all edges to obtain undirected graphs. The average degree often increases when we do this. 

For random graphs, we build each graph at different scales, which represents a generation with $2^\text{scale}$ vertices.  Graphs are summarized in Table  \ref{tab:graph_stats}.

    \label{sec:Graphs}
\begin{table}[]
    \centering

\begin{tabular}{@{}lrrrr@{}}
\toprule 
Graph               & Vertices & \begin{tabular}{@{}r@{}}Connected \\ Vertices  \end{tabular} & \begin{tabular}{@{}r@{}}Undirected \\ Edges  \end{tabular} &  \begin{tabular}{@{}r@{}}Max \\ Degree  \end{tabular}\\
\midrule 
RMAT s21 & 2.1M & 1.3M & 64M & 107K  \\
RMAT s25 & 34M & 14M & 1.0B &  1.9M \\
Forest Fire s28 (FF) & 268M & 268M & 1.2B & 1.4K \\
RMAT s28 (RMAT) & 268M & 97.7M & 8.2B & 8.1M \\
Erdos Renyi s28 (ER) & 268M & 268M & 9.4B & 148\\ 
RMAT s33  & 8.6B & 3.5B & 265B & 42M \\
\midrule 
cit-Patents & 3.7M & 3.7M & 33M & 793  \\ 
soc-liveJournal (LJ) & 4.8M & 4.8M & 86M & 22.9K \\
com-orkut (Orkut) & 3.5M & 3.1M & 234M & 33.3K \\
Twitter (Twitter) & 61.6M & 41.7M & 2.4B & 3.1M \\
\bottomrule 
\end{tabular}

    \caption{Statistics of the Random and SNAP Graphs.}
    \label{tab:graph_stats}
\end{table}

% For the random graphs, we build multiple trials of graphs from scale 8 to 24 ($2^8$ to $2^{24}$ vertices as in Graph500), and project properties at large-scale based on log-linear extrapolation. This is accurate for all the models concerned and the projections in this space seem consistent. 
% %We use a scale 28 large graph for to collect performance on the simulator (see also Section \ref{subsubsec:modeling}). 
% Graphs are processed by splitting the high-degree vertices, rendering them easier to load balance~\cite{Gonzalez-2012-powergraph}. 

\subsection{K-truss Performance}

The performance of UpDown for K-Truss was recognized with an "Innovation Award" in the MIT graph challenge \cite{HPEC-ktruss}. The results in Table~\ref{tab:ktruss-cpu-gpu} show that a single UpDown node achieves a 53-fold and 38-fold speedup over an H200 GPU and 1238-fold and 2054-fold speedup relative to a multicore CPU. Moreover, \cite{HPEC-ktruss} shows that a set of 64 UpDown nodes outperforms the full 96k BlueGeneQ nodes on this task. 

%We also evaluate performance in terms of a \emph{real-time graph}. This reflects the size of the problem that can be solved in \emph{1 second}, \emph{100 ms}, and \emph{10 ms} as we vary computational resources. 

%We implemented and evaluated K-truss on UpDown, and as the first scalable implementation submitted, were recognized with an "Innovation Award" in the MIT graph challenge \cite{HPEC-ktruss}. 

% \begin{figure}
%     \centering
%     \includegraphics[width=0.45\linewidth]{figures/ktruss-cpu-gpu}    \includegraphics[width=0.45\linewidth]{figures/ktruss-cpu-gpu-speedup}
%     \caption{K-truss performance on cit-Patents graph (4M vertices).}
%     \label{fig:ktruss-cpu-gpu}
% \end{figure}

\begin{table}
  \centering
  \small
  \setlength{\tabcolsep}{2pt}
  \begin{tabular}{@{}lrrlrrl@{}}
    \toprule
    & \multicolumn{3}{c}{Ktruss $k=3$ time (ms)}
    & \multicolumn{3}{c}{Ktruss $k=k_{\max}$ time (ms)} \\
    \cmidrule(lr){2-4} \cmidrule(lr){5-7}
    Platform
    & Value & Rel. & 
    & Value & Rel. &  \\
    \midrule
    CPU 
      & 2613 & $1.0\times$ & \minibar{2613}{2613}
      & 5073 & $1.0\times$ & \minibar{5073}{5073} \\
    cuGraph 
      & 111 & $23.5\times$ & \minibar{111}{2613}
      & 94 & $53.9\times$ & \minibar{94}{5073} \\
    UpDown
      & 2.1 & $1238\times$ & \minibar{2.1}{2613}
      & 2.5 & $2054\times$ & \minibar{2.5}{5073} \\
    \bottomrule
  \end{tabular}
  \caption{K-truss performance on cit-Patents graph (4M vertices). Relative values are versus CPU (AMD EPYC 9754),
  cuGraph (H200), and UpDown (1-node).}
  \label{tab:ktruss-cpu-gpu}
\end{table}

%\aac{performance comparison to CPU and GPU graphs}

%\aac{scaling graph}

% \begin{figure}
%     \centering
%     \includegraphics[width=0.75\linewidth]{figures/ktruss-realtime}
%     \caption{K-truss real-time analysis, various deadlines.}
%     \label{fig:ktruss-realtime}
% \end{figure}

\subsection{K-core Performance}

%The k-core algorithm identifies subgraphs where each vertex has degree at least $k$~\cite{kcore-paper}. The implementation for UpDown successively removes vertices of degree smaller than $k$ via load balanced graph computations using~\cite{HPEC-lb}. This involves a sequential iteration through all values of $k$. \jiya{Should we remove this paragraph because it is already in section A?}

% \begin{figure}
%     \centering
% \includegraphics[width=0.45\linewidth]{figures/kcores-cpu-gpu}    \includegraphics[width=0.45\linewidth]{figures/kcores-cpu-gpu-speedup}
%     \caption{K-cores performance on an RMAT scale 21 graph. UpDown significantly outperforms an H200 GPU, both outperform the CPU significantly.}
%     \label{fig:kcores-cpu-gpu}
% \end{figure}

\begin{table}
  \centering
  \small
  \setlength{\tabcolsep}{2pt}
  \begin{tabular}{@{}lrrlrrl@{}}
    \toprule
    & \multicolumn{3}{c}{K-Core time (ms)}
    & \multicolumn{3}{c}{K-Core GTEPS} \\
    \cmidrule(lr){2-4} \cmidrule(lr){5-7}
    Platform
    & Value & Rel. &
    & Value & Rel. & \\
    \midrule
    CPU
      & 1428 & $1.0\times$ & \minibar{1428.5}{1428.5}
      & 0.66 & $1.0\times$ & \minibar{0.666}{91.72} \\
    cuGraph
      & 389 & $3.67\times$ & \minibar{389.6187}{1428.5}
      & 2.44 & $3.67\times$ & \minibar{2.443}{91.72} \\
    UpDown
      & 10 & $138\times$ & \minibar{10.38}{1428.5}
      & 95.5 & $138\times$ & \minibar{91.72}{91.72} \\
    \bottomrule
  \end{tabular}
  \caption{K-Core performance on RMAT 21. Relative values are versus GBBS on CPU (AMD EPYC 9754), cuGraph (H200), and UpDown (1-node).}
 
  \label{tab:kcore-cpu-gpu}
\end{table}

\begin{figure}
    \centering
\includegraphics[width=0.85\linewidth]{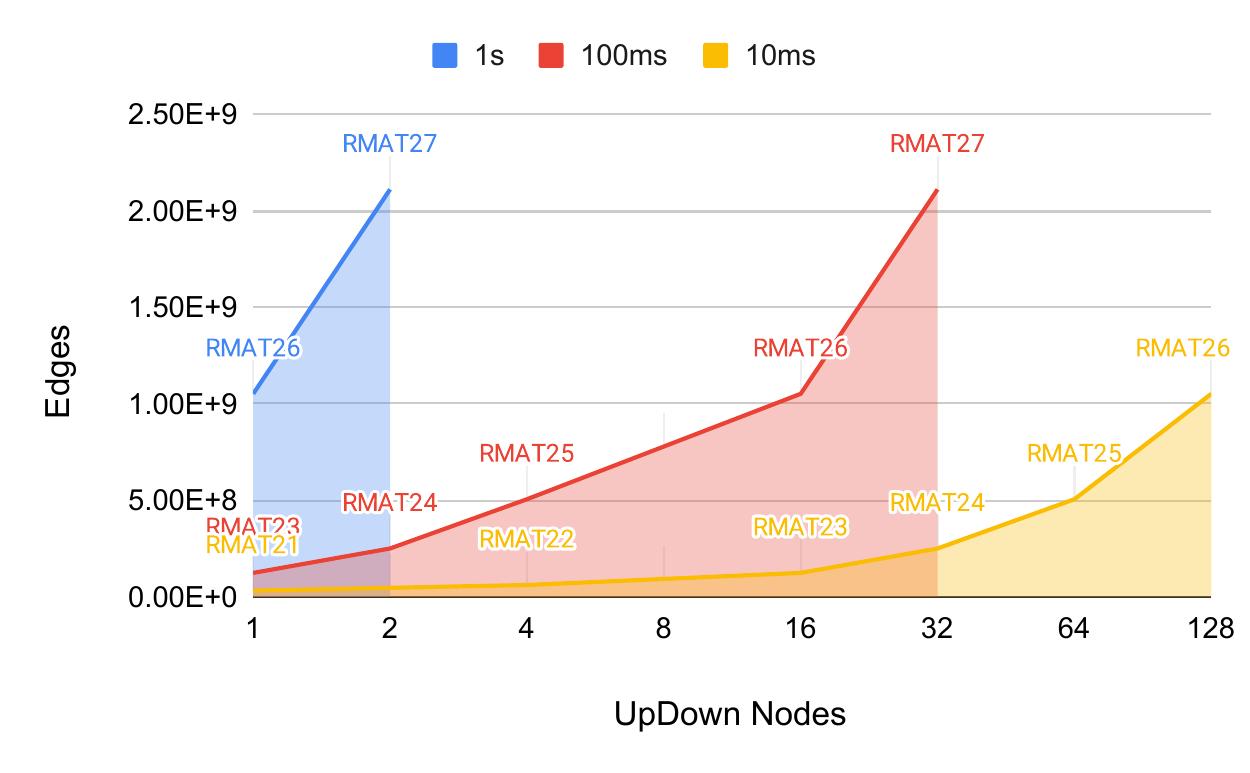}
\vspace*{-0.1in}
    \caption{K-Core real-time analysis, various deadlines.}
    \label{fig:kcores-realtime}
\end{figure}

A single UpDown node outperforms the CPU by 138-fold, and the H200 GPU by 38-fold, as shown in Table~\ref{tab:kcore-cpu-gpu}. A more interesting result is what we call a \emph{real-time} graph (see Figure~\ref{fig:kcores-realtime}). This shows the size of the graph that can be analyzed for a given deadline of 10ms, 100ms, and 1 second.  For example, UpDown can meet a 10ms deadline for an RMAT-26 graph with 128 nodes.  A 100ms deadline can be achieved with only 16 nodes, and a 1s deadline with only 1 node. For an RMAT-27 graph, slightly larger UpDown systems are required.

%\aac{performance comparison to CPU and GPU graphs}

%\aac{scaling graph}

\subsection{Breadth-first Search (BFS)}
We study push-based BFS and push-pull-based BFS~\cite{Beamer2012} in detail, as they are common graph system benchmarks. 
%The latter algorithm reduces work in the pull phase when most of the vertices have been visited. 
%In each BFS round, we have a frontier of newly-visited vertices. We mark the neighbors of these vertices with their distance, but only if they have not yet been visited, and add them to the next frontier.

% \aac{redundant?}
% For a push phase, each vertex in the frontier writes a new distance to its neighbors that have not yet been marked, and adds them to the next frontier.  During a pull phase, each unmarked vertex in the graph 
% %first checks if it is already done, and if not, 
% checks its neighbors and, if necessary, adds itself to the next frontier. 

% \aac{tighten these two paragraphs}
% We describe how to map push BFS onto the \SysName fine-grained architecture.  First, define a basic \texttt{parallel\_for(start,end,F)} primitive for parallelism on an interval.  This can be done divide-and-conquer: while \text{start != end} launch two \texttt{parallel\_for()} tasks, on the first and second half of the range.  If the interval is size one, execute \texttt{F(start)}.

For push BFS on \SysName, in each iteration, we run a \texttt{parallel\_for} over the vertices in the frontier. Each vertex also uses a second \texttt{parallel\_for} to send an update to each of its outgoing neighbors. All the updates to a vertex will be sent to the same lane to avoid race conditions. The receiver checks if the vertex has been visited. If not, it sets itself and adds to the next frontier. For scalability, the frontier is distributed across nodes, and a group of processing \lanes manages a subset of the frontier (i.e., reads from the old frontier and inserts vertices to the new frontier) locally. 
% \todo{Ivy, I'm not sure how the frontier is managed in your approach} 
The program terminates when all local frontiers are empty.

In the push-pull algorithm, we run push BFS for a few steps and then switch to a pull-based algorithm following~\cite{Beamer2012}. For pull BFS, we perform a \texttt{parallel\_for} over all vertices. Each vertex first checks whether it has already been visited; if so, it is done.  If not, the vertex performs a \texttt{parallel\_for} over its neighbors, checking if any are in the frontier.  If it finds a neighbor on the frontier, it marks the vertex as visited and adds it to the next frontier.
%\footnote{The frontier can be tracked by using the distances; when the distance is equal to the iteration number, the vertex is in the frontier.} 
Switching between push and pull is based on frontier size, using a pull-based approach when the frontier is large to eliminate redundant update messages.

To handle skew, we split high-degree vertices into multiple siblings if their degree exceeds 512. 
%In BFS, it is simpler to handle the split vertices compared to PageRank,
Because all the sibling vertices belonging to the same original vertex are added to the frontier when one of the sibling vertices is touched. Therefore, when a vertex is touched (i.e., added to the frontier), all its siblings are also added to the frontier during the BFS iteration.

%In addition to the naive approach, we also present 
Another approach to handle skew is the load-balanced push BFS (LB Push BFS).  This algorithm relies on a self-balancing variant of \texttt{parallel\_for} and demonstrates the programmability of the fine-grain architecture.  The primary advantage of the load-balanced BFS is that it can process skewed graphs without vertex splitting.  The key idea is to track the amount of work in each parallel region and assign a proportional number of workers to it.  Each recursive call to the \texttt{load\_balanced\_parallel\_for} is responsible for mapping a portion of the computation over a specific set of threads.  At each step of the recursion, we estimate the work fraction corresponding to each half of the indices and assign a proportional fraction of workers to that half of the \texttt{load\_balanced\_parallel\_for}. For BFS, this amount of work is a fraction of the edges in that sub-interval (half in this case), and this fraction is tracked in each iteration. This uses the load-balancing tool described in prior work~\cite{HPEC-lb} and does not use the vertex splitting on the graphs, unlike the other methods.

\subsection{PageRank (PR)}

% Table \ref{tab:algorithm} summarizes the variants of the PageRank and BFS algorithms  evaluated. % in the paper.
% We describe algorithms as \emph{push and pull}. \emph{Push}-variants algorithms \emph{write to neighbors}, whereas \emph{pull} variants read from neighbors. Most are implemented in the UDWeave programming language using a MapReduce-like library specialized to the UpDown architecture~\cite{Fell2025}. This enables simple operations like \verb|parallel_for| over all vertices and the operations can have side effects since the system is a global shared memory architecture. 

%Depending on the graph data and the network topology, the right choice can improve performance~\cite{whang2015scalable}.

% of two PageRank algorithms (push and data-driven PageRank) and three BFS algorithms (push, push-pull, and load-balanced BFS) using simulation and project the performance for all five but push-push BFS in a full-scale \SysName system. \ivy{table consumes too much space, switch to text description. }

PageRank (PR) calculates the importance of a vertex by weighing how many important vertices are connected to it.  Each vertex shares some of its current importance with its neighbors in each iteration.  PR can be implemented by a simple \texttt{parallel\_for} over all vertices. In each iteration, each vertex sends a fraction of its current score to all its neighbors. Each vertex receives them, sums them up, and also adds a small constant offset.  %up all of the scores it receives in each iteration.  
PageRank terminates when all updates have a value less than $1/|V|$. This gives a slightly growing iteration count for each problem. 
%We use GTEPS (giga-traversed edges per second) as the performance metric for PR since it reflects the updates pushed or pulled along edges each second. 
% As a pragmatic matter, this is the number of graph edges divided by iteration time. 

%\ivy{add data-driven pagerank algorithm}
One downside of the Push PageRank algorithm is that convergence is non-uniform --  some vertices take more iterations, as noted by McSherry~\cite{mcsherry2005-uniform}.  A recent refinement and formal algorithm that exploits this is the data-driven PageRank algorithm~\cite{whang2015scalable}.  It identifies a list of vertices that have changed enough to cause other nearby vertices to \emph{possibly} violate their convergence tolerance. This can be detected simply based on the magnitude of the update. This defines the active set, and only that set sends updates each iteration. 
%The algorithm is given in Algorithm~\ref{alg:ddpr}. 
The active set is encoded with a bitmask over all vertices, and used to control the \verb#parallel_for#.   This restricts PR value propagation to local subgraphs, reducing the work per iteration. 

Except for LB Push BFS, all experiments are done with vertices split to a maximum degree of 512 to balance the load distribution across the system. However, this approach incurs overhead for merging the result back into the original vertex's PageRank value. For push PageRank, the merging is done within the iteration: when the vertex program is launched in a lane, it checks if the vertex is a split vertex, and if so, it will read the value of the previous iteration from all its sibling vertices to calculate the value for the original vertex before pushing the value to its neighbors. In data-driven PR, the value is merged with an additional iteration: since the active set is updated after all sibling vertices have computed their partial values from their neighbors, this update must be synchronized across all split vertices. 

In experiments with different algorithms, we found that data-driven PageRank converges in a single iteration on ER graphs. This is expected because all nodes in an ER graph are approximately equivalent, and there are no vertices that attract a high rank. For the other graphs, we run the first two iterations of data-driven PageRank without tracking the working set, since we found we needed a few iterations to get close to a solution before tracking it.

\subsection{PageRank and BFS Simulation Results}
\label{sec:simulations}

% \begin{table*}[t]
%     \centering
%     % \resizebox{\columnwidth}{!}{ 
%     \input{tables/algorithm}
%     % }
%     \caption{PageRank and Breadth-first Search variants. }
%     \label{tab:algorithm}
% \end{table*}

%\aac{Ivy update graphs}
% \todo{redo with new baseline of just the 1 node numbers or something}
%In these experiments, we
We evaluate the performance of the five variants of PageRank and BFS algorithms on \SysName, examining performance scaling from 1 to 256 nodes (524,288 processing \lanesnospace). Speedups are self-normalized, and the runtimes are used to compute GTEPS. 
Effective GTEPS reflect the push PR computation work, but with the time from data-driven PR. 
%used to evaluate variants of the algorithms on the \SysName system. 
% Note that we self-normalized the performance to 1 node performance of the same algorithm (i.e., .
% h, we use \emph{effective GPEPS}, which gives the standard algorithm the effective computation rate to converge to the same tolerance, as the metric. 
% \ivy{remove GPEPS change to GTEPS and make a table for GTEPS and effective GTEPS}
% compare the \SysName performance to a 1-core CPU baseline program implemented in C++. The goal is to establish the relative performance possible with a system of \SysName cores compared with a single CPU core.  \brian{what this baseline is needs to be defined}
% \begin{table}[]
%     \centering
%     %\resizebox{\columnwidth}{!}{ \input{tables/eval_metrics}
%     %}
%     \footnotesize
%     \input{tables/eval_metrics}
%     \caption{Performance metrics for PR and BFS. % to evaluate PR and BFS performance on \SysName system.
%     }\label{tab:eval_metrics}
% \end{table}

\subsubsection{BFS}

% \begin{figure*}[ht]
% \centering
%   \includegraphics[width=0.49\linewidth]{figures/BFS_speedup.pdf}
%   \includegraphics[width=0.49\linewidth]{figures/BFS_GEPS.pdf}
%       \caption{\pname BFS Performance.}
%   \label{fig:BFS}
%   \vspace{-0.5cm}
% \end{figure*}

\begin{figure*}[ht]
\centering
  \includegraphics[height=0.17\linewidth]{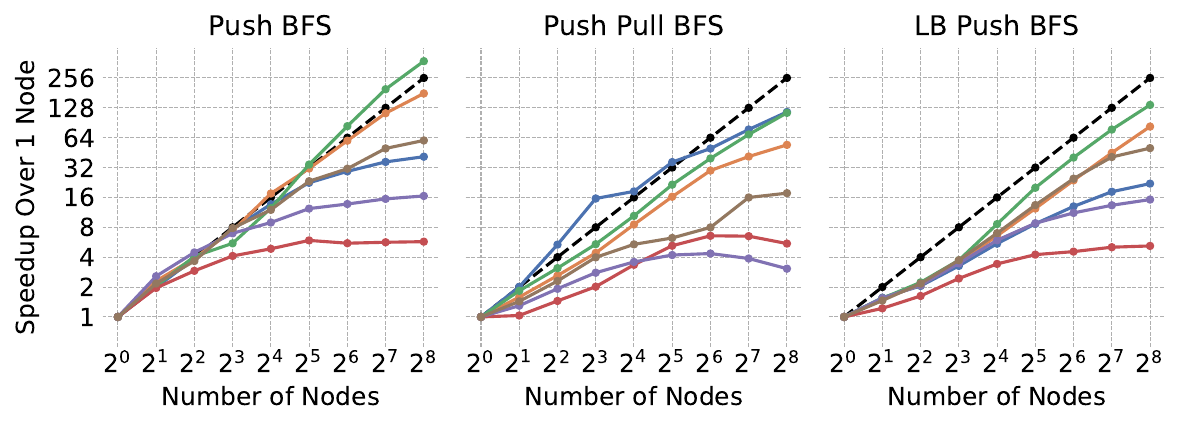}
  \includegraphics[height=0.17\linewidth]{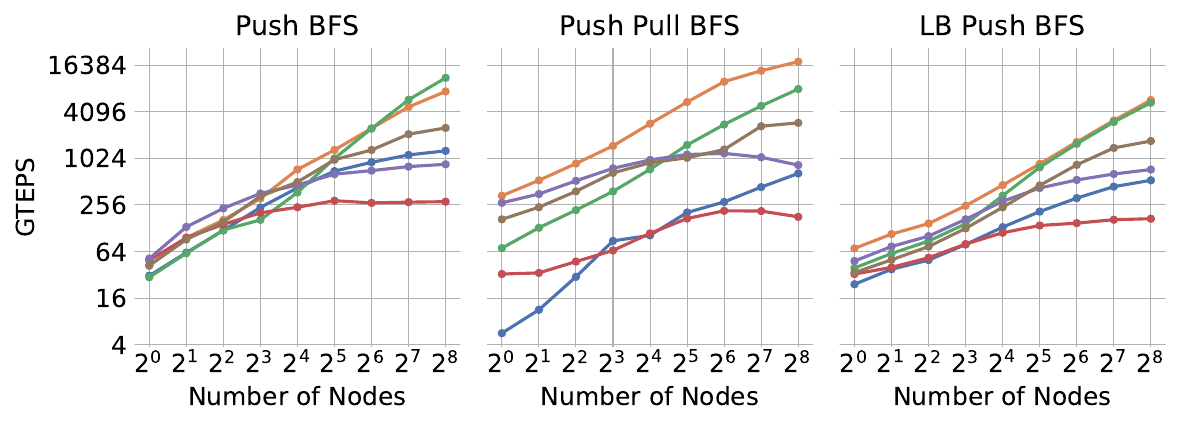}
      \caption{Left: Speedup over one node performance on \pname across various graphs for push, push-pull, and load-balanced (LB) push BFS, respectively. Right: GTEPS for all three BFS algorithms on 1 to 256 nodes \SysName system. }
  \label{fig:bfs-perf}
\end{figure*}

Figure~\ref{fig:bfs-perf} (left) shows \SysName's push, push-pull, and load-balanced BFS speedup and GTEPS for 1 to 256 nodes.  
For scaling, push BFS achieves 16x to 377x, with the larger RMAT and ER graphs achieving 178x and 377x.  Push BFS achieves scalable performance on a variety of skewed graphs by leveraging vertex splitting for good load balancing and efficient dynamic parallelism on individual vertices and edges.  In absolute performance (Figure \ref{fig:bfs-perf} (left)), 
push BFS achieves a maximum of 11,255 GTEPS for ER on 256 nodes. The average across the graphs on 256 nodes is 3,952 GTEPS.  

Compared to the push BFS, the push-pull BFS switches to the pull phase when the frontier size is large to eliminate redundant updates to the same vertex from multiple neighbor vertices in the frontier. 
As expected, this is faster and produces a maximum of 18,231 GTEPS on the RMAT graph and an average GTEPS of 5,154 across the graphs. 
Compared to single-node performance, push-pull BFS produces a speedup of up to 116x for ER (116x) and Forest Fire graphs (113x) on a 256-node machine. The speedup is only 3x on Orkut because the graph is too small.
%i.e., too little work to be distributed to 256 nodes. 

We highlight that the load-balanced BFS achieves a high computation rate on graphs as skewed as RMAT without any preprocessing or vertex splitting via an optimized \texttt{parallel\_for} primitive.  Load-balanced push BFS achieves speedups ranging from 15x to 137x on 256 nodes compared to single-node performance, with ER achieving the highest speedup.  It achieves a maximum of 5,824 GTEPS on RMAT with 256 \SysName nodes and an average of 2,391 GTEPS across the six graphs.  
 This shows that applications can effectively use \pnamenospace's hardware mechanisms for fine-grained parallelism to spread work with moderate software management, showcasing the programmability of \SysName and the potential of its fine-grained parallelism.

% \ivy{report the absolute performance }
% One \SysName accelerator with 64 cores delivers 45x improvement on average across the 8 graphs studied. With hardware parallelism of 4,096 cores, \SysName accelerators achieve up to 3,147x improvement and an average of 2,682x. When comparing the performance of 64 cores to that of 4,096 cores, with 64 times more cores, \SysName's push BFS program runs 235x faster.

% \begin{figure}[h]
%     \centering
%     \includegraphics[width=0.5\textwidth]{figures/Push BFS Scaling.png}
%     \caption{Push BFS Speedup (1-4,096 cores), for Scale 21 RMAT, ER, Forest Fire, and SNAP graphs, normalized to 1 core CPU }
%     \label{fig:bfs-speedup}
% \end{figure}
% The performance speedup of the push-pull BFS (refer to Algorithm \ref{alg:pushpullbfs}) over the push CPU BFS program is shown in Figure \ref{fig:push-pull-bfs-speedup}.

% \ivy{Brian, can you add two sentences that talk about how your approach is better/different from the push-based approach in terms of performance?} {FYI the summarized performances are in the google drive {https://docs.google.com/spreadsheets/d/1Y3rimSlL_OtxptHxNXnD5qVjF6QQVVyL5GPcWv-isUQ/edit?gid=1515426037#gid=1515426037.}

% \begin{figure}[h]
%     \centering
%     \includegraphics[width=0.5\textwidth]{figures/Push Pull BFS Scaling.png}
%     \caption{Push-Pull BFS Speedup (1-4,096 cores), for Scale 21 RMAT, ER, Forest Fire, and SNAP graphs, normalized to 1 core CPU }
%     \label{fig:push-pull-bfs-speedup}
% \end{figure}

%performance

\subsubsection{PageRank}

\begin{figure*}[ht]
~\hfill 
  \includegraphics[height=1.25in]{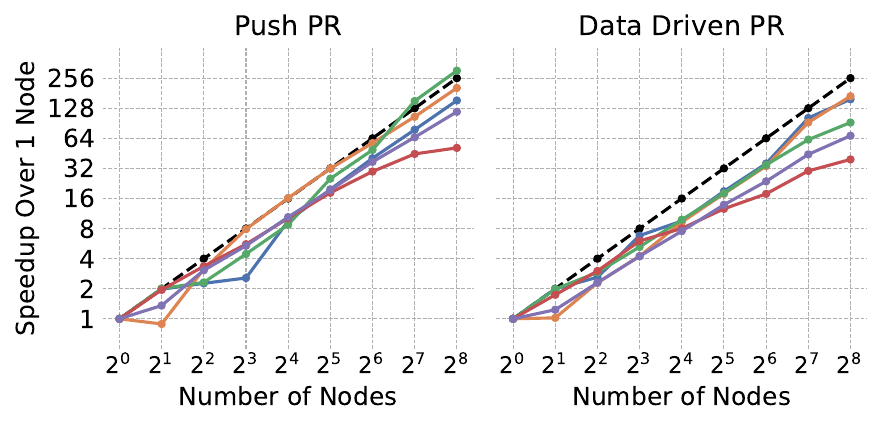}
  \hfill 
  \includegraphics[height=1.25in]{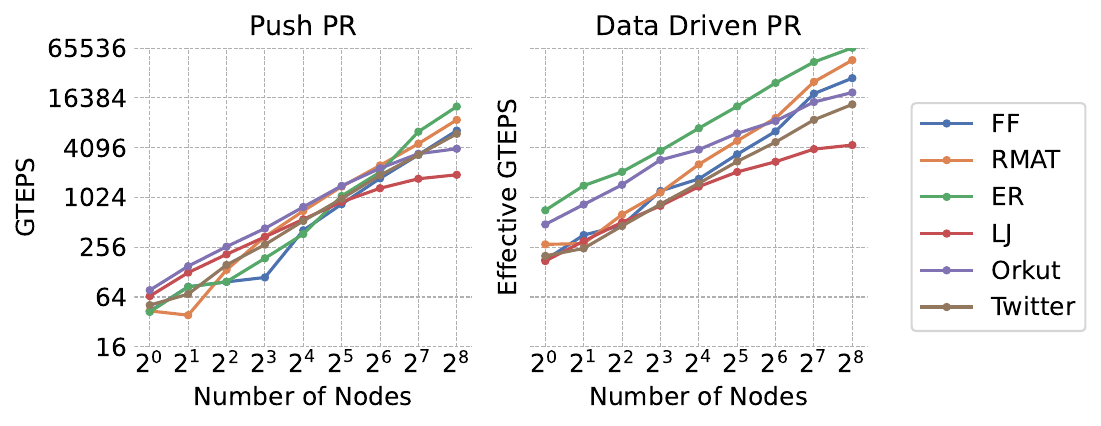}
  \hfill~ 
      \caption{Left: Speedup over one node performance on \pname across various graphs for push and data-driven PR, respectively. Right: GTEPS for push and effective GTEPS for data-driven PR algorithms on 1 to 256 nodes \SysName system.}
  \label{fig:pr_perf}
\end{figure*}

% remove sub title, exchange solid and dashed line

% \begin{figure}[h]
%     \centering    \includegraphics[width=0.5\textwidth]{figures/Push PageRank Scaling.png}
%     \caption{Push PageRank Speedup (1-4,096 cores) for Scale 21 RMAT, ER, Forest Fire, and SNAP graphs, normalized to 1 core CPU }
%     \label{fig:pr-speedup}
% \end{figure}

Figure~\ref{fig:pr_perf} (left) shows speedup and (right) GTEPS for push and data-driven PR.  Speedups are normalized to 1-node performance. 
As for scaling, the push PR shows speedups ranging from 51x to 304x, with the highest speedups on RMAT and ER.
% For simplicity, we ran the data-driven PR for up to 5 iterations instead of to convergence and computed effective GTEPS accordingly. 
For push PR, \SysName achieves a maximum of 12,850 GTEPS on the ER graph at 256 nodes and an average of 7,660 GTEPS across the five graphs studied.
% As for scaling, the push PR shows a maximum 304x speedup at 256 nodes, with an average improvement of 166x across graphs.  
% \jiya{

% }
Performance is affected by the ratio of vertices per processing \lane due to a software cache implemented in the lane scratchpad memory that combines push updates.
When the vertices do not fit in the cache, performance drops, but increases as the machine size scales up.
%: a higher ratio leads to worse performance than expected because the atomic updates to merge the push updates are done via a software cache implemented with the scratchpad. When the number of vertices per \lane exceeds the scratchpad capacity, the cache conflicts would increase, slowing down the progress. 
Orkut performance scales poorly beyond 32 nodes because the graph is small.  %enough to generate enough work to saturate the system beyond 32 nodes. For 
Larger graphs (e.g., RMAT-s28) achieve near-perfect scaling (black dashed line). 
% At 64 MIMD cores (1 accelerator), 107-fold  improvement is achieved. At 2,048 cores (1 \SysName node), speedups range from 3,007x (cit-Patent graph) to 4,666x (LiveJournal).  For two nodes, speedups reach 5,495-fold and an average of 2,257x improvement across graphs.\footnote{Two \SysName nodes consume similar power to a single NVIDIA H100 GPU.} 

% \jiya{Should we point out that the reason why Lj and Orkut cannot scale well on 256 nodes is that their graphs are too small, which makes the workload per core too small and the synchronization overhead occupies most of the execution time?}

% \begin{figure}[h]
%     \centering
%     \includegraphics[width=0.5\textwidth]{figures/Data-Driven PageRank Scaling.png}
%     \caption{Data-driven PageRank Speedup (1-4,096  cores), relative to push-based PageRank baseline, for Scale 21 RMAT, ER, Forest Fire, and SNAP graphs, normalized to 1 core CPU }
%     \label{fig:dd-pr-speedup}
% \end{figure}

% We measure the improved algorithm, Data-driven PageRank (see Algorithm \ref{alg:ddpr}). 
% The performance for data-driven PageRank is plotted in Figure \ref{fig:dd-pr-speedup}.
%Compared to the push PR
Data-driven PR reduces the work by intelligently updating only a subset of the graph.
%at later iterations. It is designed to match the convergence criteria of push PR, and it provides the same guarantee on the output. 
It achieves speedups ranging from 39x to 169x, with the highest speedups on RMAT and Forest Fire.  
% a maximum effective GTEPS of 65,977 for ER on 256 nodes and an average of 34,761 across the five graphs. 

% giving the best.
% This leads to a 93\% improvement over push PageRank on 512 nodes \SysName. 
As for scaling performance, on 256 nodes, data-driven PageRank delivers an average 105x increase in edges processed per second compared to the 1-node run. 
On average, data-driven PR's performance speedup over push-based PR ranges from 2.3x to 5.3x, with RMAT and ER giving the best results, as they eliminate the cache effects described for push PR, reduce work after the first few iterations, and converge faster.  
%the need for atomic updates using software cache and the resulting cache effect.
% producing an average 3,150-fold speedup over the CPU baseline, and up to 7,407x in the case of cit-Patent. 

% \textbf{Summary}
% The \pname architecture is remarkably efficient in exploiting fine-grained parallelism at the vertex and edge levels.  The performance of PageRank and BFS scales to 256 nodes and achieves ten thousand GTEPS performance even with extremely skewed degree distributions (e.g., in RMAT).

\subsection{Overall Summary and Discussion}

The results show the broad capability of UpDown to accelerate graph computation.
%\subsection{Discussion}
\pname speedups exceeding 100-fold vs a CPU can shift significant analyses from batch to interactive/exploratory, reducing hours-long runtimes to minutes. The results in Table~\ref{tab:speedup-h200} summarize speedups
%across a variety of graph algorithms.  The geomean 1-node UpDown speedups are 25x geomean relative to an H200 GPU.    
relative to H200 GPUs, the performance advantage is  large, varying from 7x to 50x with a geomean of 28x. %

%\david{This last line is inconsistent with the first...}
%Overall, performance scales well for thes because \SysName effectively exploits fine-grained vertex- and edge-level parallelism, leveraging fast hardware messaging and short threads as described in Section \ref{sec:approach}. 

%, providing the potential for a much lower power 8-node UpDown system to provide a similar transformation relative to a single GPU.

% \aac{move later?}
% With scaleup of UpDown systems, additional speedups of 100,000-fold can be achieved (see Section \ref{sec:evaluation}), making analyses that are barely feasible to date (1,000's of CPU hours), achievable in seconds.  

\begin{table}
  \centering
  \small
  \setlength{\tabcolsep}{3pt}
  \begin{tabular}{rrl}
    \toprule
    Algorithm & \multicolumn{2}{c}{Speedup vs. H200} \\
    \midrule
    k-truss (k=3)  & 53   & \minibar{53}{53} \\
    k-truss (k=kmax)  & 38   & \minibar{38}{53} \\
    k-core         & 38   & \minibar{38}{53} \\
    triangle count & 8.5  & \minibar{8.5}{53} \\
    PageRank       & 44   & \minibar{44}{53} \\
    BFS            & 40   & \minibar{40}{53} \\
    CC             & 6.9  & \minibar{6.9}{53} \\
    SCC            & *    & Failed on H200 (OOM),\\
                   &      & dense algorithm used\\
    Louvain        & 42   & \minibar{42}{53} \\
    \addlinespace[1pt]
    \cmidrule(l){1-3}
    geomean        & 28   & \minibar{25}{53} \\
    \bottomrule
  \end{tabular}
  \caption{Speedups relative to H200 GPU/cuGraph.}
  \label{tab:speedup-h200}
\end{table}

\section{Projections to Full System Scale}
\label{sec:evaluation}

% We describe the methodology and experiments to evaluate the \pname system.  Extensive simulations  
% %characterize \pnamenospace's efficiency in delivering performance, 
% and then projections characterize performance at full system scale.
% %, showing the ability to achieve high absolute performance.  
% Finally, we use a network model to study   communication limits. 
% %requirements, assessing the \pname system network.
% %ability to support them.

%compare \pnamenospace's %\brian{estimated or projected?} 
%projected full-system performance with that of other scalable systems. %high-performance system results.

%\subsection{Methodology}
%\label{sec:Methodology}

In this section, we build an analytical projection model that enables us to study graph sizes that would be extremely expensive to simulate. We then validate a few outcomes from the analytical model using simulations on a supercomputer.

\subsection{Projection Methodology}

We note that the graphs analyzed in the previous section are rather small for a collection of 256 nodes. 
%We have at most 10 billion edges, which, spread over 256 nodes, is about 40 million edges \emph{per node} or 320 megabytes of graph data per node (assuming 8 bytes stored for each edge). The goals of the AGILE program that helped to develop UpDown are to understand the performance of the UpDown machine on much larger graphs that are difficult or impossible to simulate exactly. 
To address this, we used the projection methodology from the US Government Agile Program. %In this section, we explain how we estimate performance for a scale 40 RMAT graphs (1 trillion vertices with around 30 trillion edges) on the full system of 16,384 nodes. 
When projecting performance for PageRank and BFS on large synthetic graphs, we base our projections on measurements from smaller-scale graphs (s8–s24). The workload is primarily based on the total number of edges \emph{touched} by the algorithm.   In BFS, this is the expected number of frontiers along with the vertices and edges in the graph. In PR, this is the sum of degrees (also known as volumes) of updated vertices, along with the expected maximum degree to estimate the performance (to handle synchronization induced by the impact of split vertices).  For all, we include expected synchronization costs in two ways: (i) those included in the Fastsim2 measurements used to calibrate projections and (ii) through explicit modeling of tree reductions across system nodes.  Projection results are shown in GTEPS; except for data-driven PageRank, we report effective GTEPS, computed using the work-reduced time but with the original algorithm's work. 

\begin{figure}[h]
    \centering
    \includegraphics[width=\linewidth]{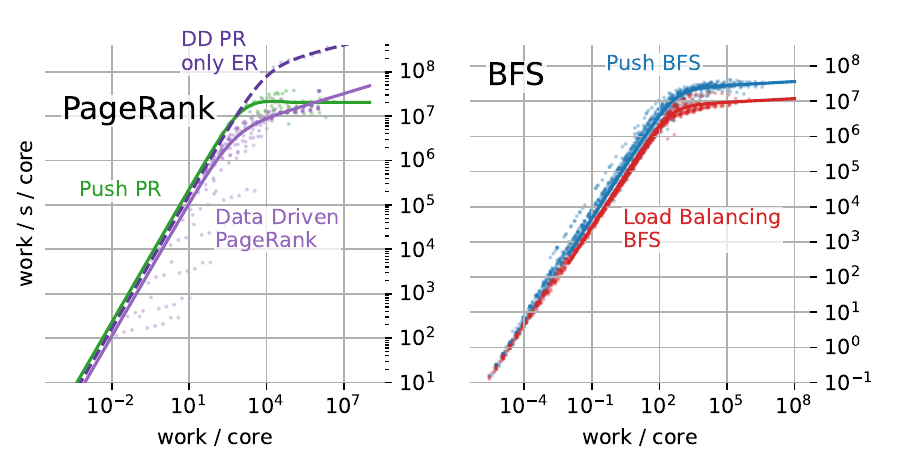}
    \caption{Per \lane work to per \lane work rate for PR (left) and BFS (right).  Data measured from the simulation are plotted as scatter points, and the fit rates used in the models are shown as lines. These show good, predictive agreement between the analytical models (lines) and data. We found that DD-PR behaved differently for ER graphs and fit a separate model.}
    \label{fig:work_rate}
\end{figure}

% To estimate the rate at which the algorithms can process the work that's assigned to each \lanenospace, we fit a modified sigmoid function $f(x\texttt{;}c,x_0,k) = \frac{c\cdot x}{(1 + (\frac{x}{x_0})^k)}$
% to fit the trend of when the serial work dominates over the communication of the machine. When enough work is assigned to a \lanenospace, we expect that the runtime will grow linearly. Figure~\ref{fig:work_rate} reports a scatter plot of the work/\lane vs. the work/\lanenospace/s of our experiments alongside with the fitted regression that we use for our models.  For Push PageRank, we use the number edges in the graph and for data-driven PageRank we use the active set volume and number of active vertices in the first 5 iterations as the work. For BFS, we measure the time to process the $i$th frontier and the number of vertices and edges iterated to build the next frontier as the work. 
% The key property of our regression is that at a certain amount of work assigned to each \lanenospace, the machine's processing rate flattens out. When a \lane doesn't have enough load then the processing rate slows down. Regression fits are not perfect, and so we apply a max cutoff to the fastest work rate per \lane observed across that algorithm's experiments (which can be seen at $10^8$ of the data-driven PageRank regression).

\begin{figure*}[t]
    \centering

    \begin{minipage}{0.49\textwidth}
        \centering   
        \includegraphics[width=\linewidth]{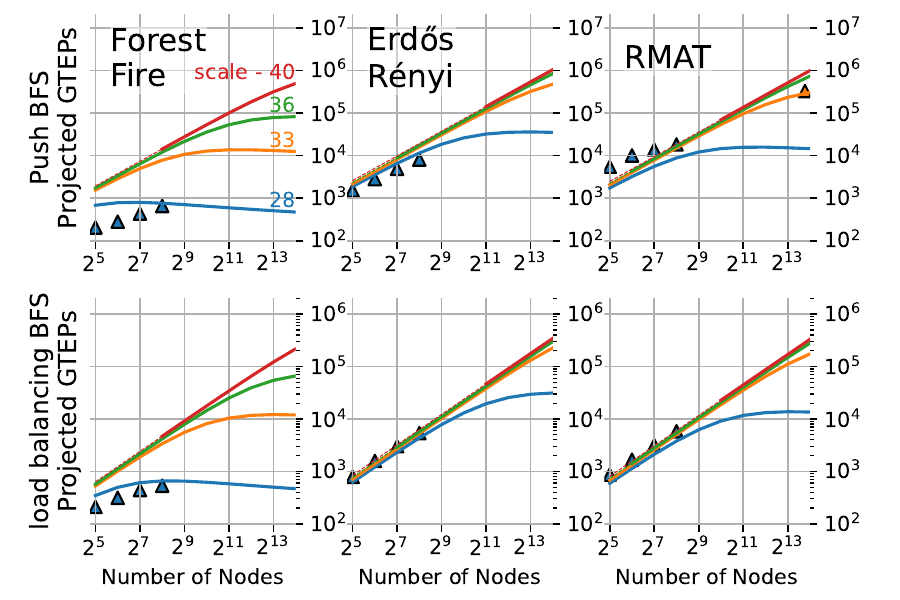}
        \caption{BFS GTEPS (64k-33M \lanes for 32-16k nodes), Scales 28-40, Forest Fire, ER, and RMAT. Solid lines are feasible. Dashed lines are projections for machines that lack sufficient DRAM to actually run the computation. The triangle markers show validation runs, see Section~\ref{sec:validation}.}
        \label{fig:bfs-speedup-projected}
    \end{minipage}
    \hfill
    \begin{minipage}{0.49\textwidth}
        \centering
        \includegraphics[width=\linewidth]{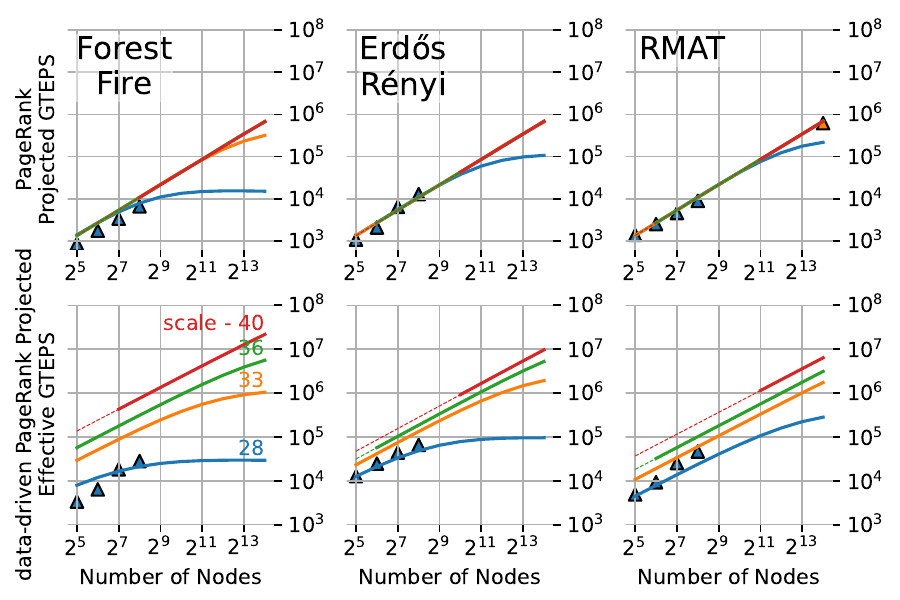}
        \caption{PageRank GTEPS (64k-33M \lanes for 32-16k nodes), Scales 28-40, Forest Fire, ER, and RMAT. Dashed lines correspond to projections on machines that do not have enough DRAM to store the edge list and the minimum number of vectors needed for the algorithm; solid lines are feasible. The triangle markers show validation runs, see Section~\ref{sec:validation}.}
        \label{fig:pr-speedup-projected}
    \end{minipage}

\end{figure*}

We calibrate projections with a computation rate based on the amount of work assigned to a lane.  Figure~\ref{fig:work_rate} reports a scatter plot of the work/\lane vs. the work/\lanenospace/s measured in experiments.   For BFS, we use work (\# of vertices and edges iterated to build the next frontier) and time (time to process the $i^\text{th}$ frontier). For Push PR, we use work (graph edges), and for data-driven PR, we use work (active set volume and \# of active vertices). Crucially, the Figure shows a reliable pattern for all simulations.  

We fit the parameters $c, x_0, k$ for a modified sigmoid function $f(x\texttt{;}c,x_0,k) = \frac{c\cdot x}{(1 + (\frac{x}{x_0})^k)}$ to the results. As the work per lane decreases, the computation rate goes down because any overheads dominate. As work per lane increases, the computation rate saturates at the maximum possible in the machine.  
The fitted functions closely resemble the measured data. We limit fits that predict indefinitely increasing processing rates by capping them at the maximum observed rate. Additionally, due to the data sparsity of PageRank for large graphs on small machines, we use the rate fit from $10^5$ work / \lane for estimates for larger loading. %We don't expect the serial work rate of a parallel algorithm to decrease as additional serial work is assigned. 
By evaluating this fitted function for the data distribution corresponding to a much larger graph, we can project performance for larger graphs.
While a single model was sufficient in most cases, data-driven PR showed different behavior on ER graphs. Consequently, we fit a separate model for ER graphs for data-driven PageRank only.

% \textbf{\textcolor{red}{XXXX-These need to be updated based on the final figure-XXXX} }
% \jiya{Dada-driven is faster both on simulated and projection.}
%  Next, we study data-driven PR
%  %what is possible by changing the algorithm in a way that should utilize the graph structure to reduce work 
%  (row 2 in Figure~\ref{fig:pr-speedup-projected}).  Data-driven PR reaches markedly higher performance, roughly 10-fold for RMAT, and 10 to 100-fold for Forest Fire. Its advantage is less on ER graphs by about 20-30\% because ER graphs only need 1 iteration at $\varepsilon=1/n$ tolerance, causing data-driven PR to traverse the same number of edges as Push PR, but with more overhead to maintain active vertex sets. 
%  %Note that these effective GTEPS rates differ from Figure~\ref{fig:pr_perf} because we project for tolerance-dependent iterations instead of 5 as in Figure~\ref{fig:pr_perf}. 
%  %We also note that a slightly outdated set of performance measurements were used in fitting the DD-PR performance model. Evidence on the difference with PageRank suggests this would not change values by more than 5\% and so we regard these are reliable. 

\subsubsection{BFS} To project performance, we use work (adding new vertices to the frontier) and time (traversing the edges outgoing from the frontier or incident to the unvisited vertices). We compute the runtime of the BFS algorithms with the number of vertices and edges expected to be traversed divided by the available \lanes and add in $\log_2(node)$ times the round-trip time to access DRAM to add another layer for frontier synchronization costs. We model the runtime for a scale $s$ graph using a $p$ node machine of 2048 \lanes with 
\begin{tightcenter}\footnotesize\begin{align*}
    &\underbrace{\bigg(\frac{2\cdot \texttt{edges(}s\texttt{)} + \texttt{vertices(}s\texttt{)}}{p\cdot 2048}\bigg)/ \texttt{work\_rate(}\dots\texttt{)}} _{\text{time traversing graph   }} +\\&\substack{\texttt{DRAM}\\ \texttt{roundtrip}} \cdot \bigg [\underbrace{\log\bigg(\frac{\texttt{max\_degree(}s\texttt{)}}{\texttt{split\_size}\cdot p}\bigg)}_{\text{split vertex reduction}} + \underbrace{2\log(p)}_{\substack{\text{iteration}\\\text{sync}}}]\bigg]\cdot \texttt{frontiers(}s\texttt{)}.
\end{align*}\end{tightcenter}

Projections in Figure~\ref{fig:bfs-speedup-projected} show that Push BFS achieves high performance on \pname across the board, with approximately 1M GTEPS for RMAT (990K) and ER graphs (1.04M), but a lower 494K GTEPS for Forest Fire, due to its larger diameter.  Load-balanced BFS has about one-third of the performance on ER and RMAT graphs, and half on Forest Fire graphs. But it also shows \pnamenospace's effective programming as the method does not require any preprocessing to achieve this performance.  % and performs comparably.
%can still perform at the same order of magnitude. 

% \begin{figure}[ht]
%     \centering   
%     % \includegraphics[width=0.5\textwidth]{figures/projections/BFS_projected_GTEPSv3.pdf}
%     \includegraphics[width=\linewidth]{figures/projections/BFS_projected_GTEPSv4.pdf}
%     \caption{BFS GTEPS (64k-33M \lanes for 32-16k nodes), Scales 28-40, RMAT, ER, and Forest Fire.   Solid lines are feasible.  Dashed lines are projections for machines that lack sufficient DRAM to actually run the computation. The triangle markers show validation runs, see Section~\ref{sec:validation}.} %to store the edge list and vertices and as well as the size of the maximum frontier size,
%     %solid lines are feasible.}
%     \label{fig:bfs-speedup-projected}
% \end{figure}

\subsubsection{PageRank} 
The amount of work in PR variants is highly predictable as graphs increase from scale 28 to 40. Data-driven PR accesses different vertices depending on the tolerance and graph topology; however, the total amount of work is well-behaved. We project the sum of volumes over all iterations required to converge (i.e., reduce the update magnitude below $\varepsilon = 1/n$).  %This scales the tolerance proportional to the graph size and 
This choice helps keep the work computed at different scales consistent relative to the graph size.  We model the per-iteration work as $|E|$ and include an internode synchronization cost to merge the split vertex values. This produces a runtime model for an $s$ scale graph over $p$ nodes with 2048 \lanes of 
\begin{tightcenter}\footnotesize\begin{align*}
    & \texttt{iterations at scale $s$}\cdot\bigg [ \underbrace{\bigg(\frac{|E|}{p\cdot 2048}\bigg) / \texttt{work\_rate(}\dots\texttt{)}} _{\text{time to process edges}} +\\&\substack{\texttt{DRAM}\\ \texttt{roundtrip}} \cdot [\underbrace{\log\bigg(\frac{\texttt{maximum degree at scale $s$}}{\texttt{split size}\cdot p}\bigg)}_{\text{split vertex reduction}} + \underbrace{\log(p)}_{\substack{\text{iteration}\\\text{sync}}}]\bigg].
\end{align*}\end{tightcenter}

Projections in Figure~\ref{fig:pr-speedup-projected} show consistent performance of around 0.7M GTEPS for push PR for the biggest system and graphs. In the same scenarios, data-driven PageRank scales from 21M \emph{effective GTEPS} on forest fire to 6.5M effective GTEPS for RMAT. Note that the scale 28 results are a small amount of work that is hard to scale to the full machine size.

\subsection{Validating the Projections}
\label{sec:validation}

These projections indicate tremendous potential for the UpDown system. To help build confidence in them, we sought to validate using a simulation of 16,384 nodes with a larger RMAT graph on a scale-out distributed simulator called Fastsim3, which can simulate the full-scale system with 16,384 \pname nodes. This requires over 28.8 TB of memory and over 64 nodes on the NERSC Perlmutter system to run. We do targeted validation on the most demanding graph, RMAT, and the largest projection configuration to validate that it fits within our computational budget. The idea is that if the projection model's output is close to our simulation result for this configuration, then the other configurations should also show reasonable agreement with the projections. As shown in Figures \ref{fig:bfs-speedup-projected} and \ref{fig:pr-speedup-projected}, the simulated performance (orange triangle) closely matches the projected performance for Push PR and Push BFS, respectively. We detail differences for Push BFS and Push PR in Table~\ref{tab:validation-pr-bfs}.  While there are differences between measurements and our projections, the goal with our projections was a simple model with better than 50\% accuracy, which all of these achieve. Indeed, we exceed the projected performance for BFS on the large-scale evaluation.

\begin{table}
\footnotesize
\begin{tabular}{@{}l l c c c c@{}}
\toprule
App & Graph & \makecell{Machine\\Size} & \makecell{Simulated\\Time (ms)} &
\makecell{Simulated\\GTEPS} & \makecell{Projected\\GTEPS} \\
\midrule
\multirow{4}{*}{Push-PR}
 & FF-s28   & 256     &   3.42    &     6,587  &   8,425    \\
 %& FF-s40   & 16,384  &   --     &    --    &    682,926    \\
 \cmidrule(lr){2-6}
 & ER-s28   & 256     &    0.73    &    12,851    &   11,102    \\
 %& ER-s40   & 16,384  &   --     &    --    &    683,782     \\
 \cmidrule(lr){2-6}
 %& RMAT-s28 & 256     &    51.27    &    8,871    &  10,909     \\
 & RMAT-s33 &  16k     &    42.25    &   407,692      &    473,927    \\
 %& RMAT-s40 & 16,384  &   --     &   --     &   684,161     \\
 %\midrule
 % \multirow{4}{*}{DD-PR}
 % & FF-s28   & 256     &   0.79     &   28,353     &   21,348     \\
 % %& FF-s40   & 16,384  &    --    &   --     &   21,485,526   \\
 % \cmidrule(lr){2-6}
 % & ER-s28   & 256     &   0.14     &   65,977     &   48,815     \\
 % %& ER-s40   & 16,384  &   --     &    --    &  9,574,518      \\
 % \cmidrule(lr){2-6}
 % & RMAT-s28 & 256     &   9.71     &    46,818    &   23,972      \\ 
 % %& RMAT-s40 & 16,384  &    --    &    --    &  6,301,973      \\
\midrule
%\end{tabular}
% \label{tab:validation-bfs-pr}
% \begin{tabular}{@{}l l c c c c@{}}
% \toprule
% App & Graph & \makecell{Machine\\Size} & \makecell{Simulated\\Time (ms)} &
%\makecell{Simulated\\GTEPS} & \makecell{Projected\\GTEPS} \\
%\midrule
\multirow{4}{*}{Push-BFS}
 & FF-s28   & 256     &   0.93    &    1,282    &    760    \\
 %& FF-s40   & 16,384  &    --    &   --     &    494,012    \\
 \cmidrule(lr){2-6}
 & ER-s28   & 256     &     0.83   &   11,255     &   11,502     \\
 %& ER-s40   & 16,384  &    --    &    --    &    1,042,746    \\
 \cmidrule(lr){2-6}
 %& RMAT-s28 & 256     &    1.10    &   7,502     &     8,780    \\
 & RMAT-s33 & 16k     &   0.84     &     315,476   &    302,673    \\
 %& RMAT-s40 & 16,384  &   --     &  --      &    989,499    \\
 %\midrule
 %\multirow{4}{*}{LB-BFS}
 % & FF-s28   & 256     &     2.22   &     533   &   659     \\
 % %& FF-s40   & 16,384  &    --    &    --    &  217,698      \\
 % \cmidrule(lr){2-6}
 % & ER-s28   & 256     &   1.75     &    5,366    &  4,343      \\
 % %& ER-s40   & 16,384  &   --     &    --    &    345,483    \\
 % \cmidrule(lr){2-6}
 % & RMAT-s28 & 256     &   1.42     &  5,824      &    3,782    \\
 % %& RMAT-s40 & 16,384  &   --     &    --    &  330,023      \\
\bottomrule
\end{tabular}
\caption{Comparison of projection on various system (256 and 16K) and graph sizes (scale 28, 33).
%systems and small graphs (scale 28) and one big graph and big system scale-33. 
The projections are all within 25\% of simulation results.}
%The goal of our projections was a simple model that gets within better than 50\% error with respect to the true performance, which this achieves.  }
\label{tab:validation-pr-bfs}

\end{table}

% Add point about why only RMAT...
% Only wanted to run the big ones to make sure the largest projections were accurate.
% 

% Move performance figures here for better format

\begin{figure*}[t]
    \centering

    \begin{subfigure}[t]{0.32\linewidth}
        \centering
        \includegraphics[width=\linewidth,height=3.7cm]{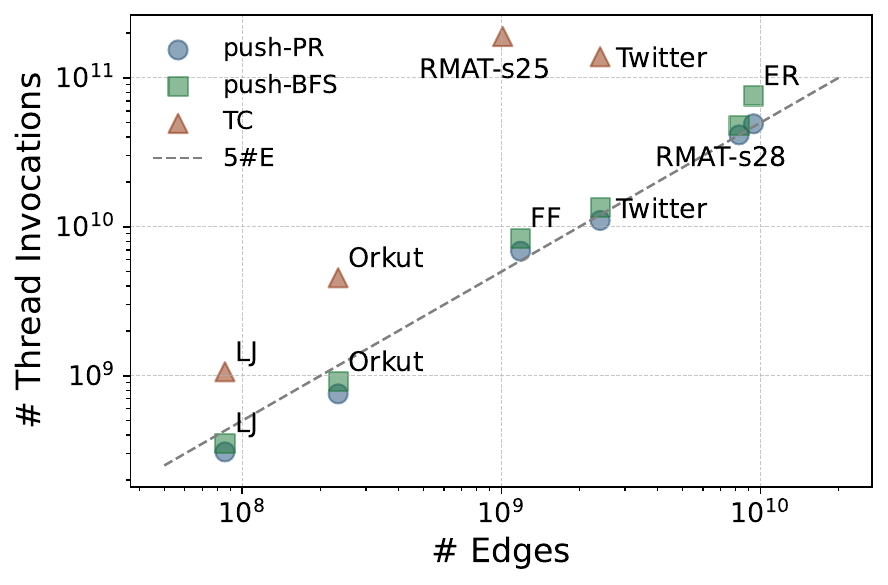}
        \caption{UpDown programs express massive parallelism: the number of thread invocations grows in proportion to graph sizes (various applications).}
        \label{fig:thread-invocations}
    \end{subfigure}
    \hfill
    \begin{subfigure}[t]{0.32\linewidth}
        \centering
        \includegraphics[width=\linewidth,height=3.7cm]{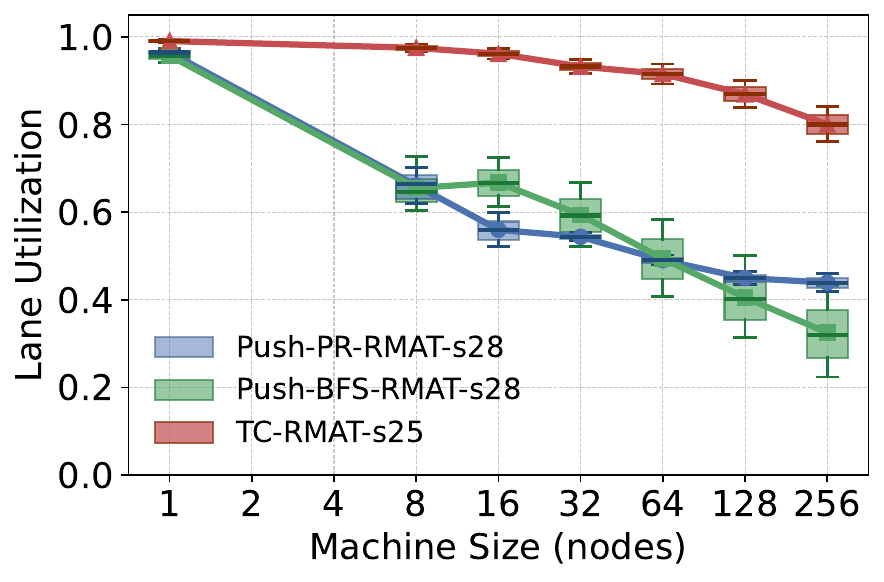}
        \caption{Average Lane utilization vs Machine size (nodes).  Whisker plots show 10th and 90th percentiles.}
        \label{fig:lane-utilization}
    \end{subfigure}
    \hfill
    \begin{subfigure}[t]{0.32\linewidth}
        \centering
        \includegraphics[width=\linewidth,height=3.7cm]{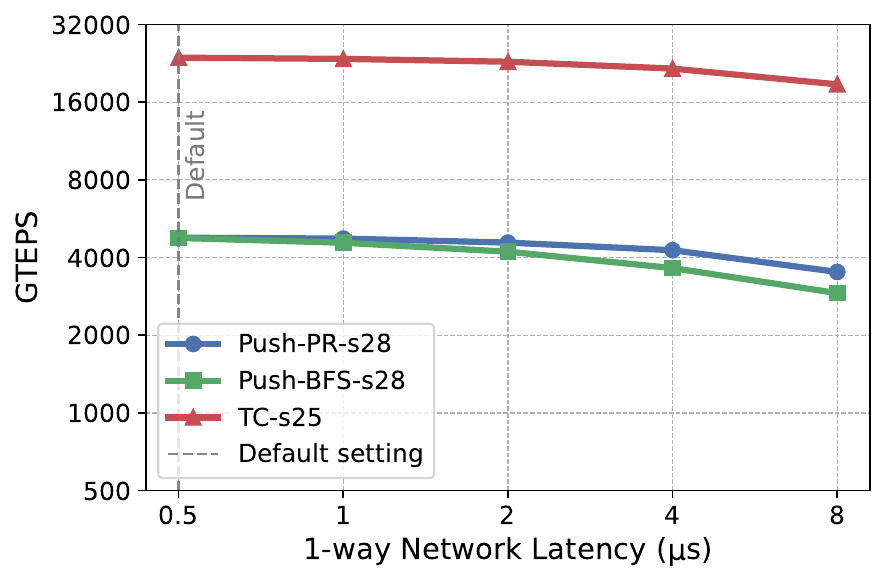}
        \caption{Application performance is robust to network latency.  A  hypothetical 16x increased latency reduces performance only marginally (128-node \pname system).}
        \label{fig:latency-tolerance}
    \end{subfigure}

    \caption{UpDown performance characteristics across applications.}
    \label{fig:three-in-one}
\end{figure*}

\subsection{Network Model}
\label{sec:network_model}

A common question about our modeled performance is how \emph{network congestion} may impact the results. 
We simulate nominal network bisection (32PB/s), injection bandwidth (4.4TB/s/node), and latency (0.5$\mu$s, 1-way), as described in Section \ref{sec:updown-architecture-approach}.
Additional studies~\cite{colley-thesis,jiya-thesis} show congestion is not a problem for networks with capabilities that are in line with co-packaged optics roadmaps \cite{Broadcom-CPO24,AyarLabs26} and recent AI announcements. These networks do represent a significant increase relative to traditional supercomputers.

\section{Understanding UpDown Performance}
\label{sec:performance-comes-from}

How can UpDown perform so well with these fine-grained algorithms?  Fundamentally, UpDown delivers scalable, high performance by maintaining high datapath utilization as an application scales from 1 to thousands of nodes.  UpDown applications are extremely parallel, creating parallelism that grows linearly with graph size.  See results in Figure \ref{fig:thread-invocations} for various applications with billion-fold parallelism that fills millions of lanes in a system, creating high utilization. 

% \begin{figure}[h]
%     \centering
%     % ~ \vspace*{1.25in}
%     \includegraphics[width=0.75\linewidth]{figures/thread_invocations_allv2.pdf}
%     \caption{UpDown programs express massive parallelism: the number of thread invocations grows in proportion to graph sizes (various applications).}
%     \label{fig:thread-invocations}
% \end{figure}

The massive fine-grained parallelism (vertex and edge) enables good machine utilization and the associated excellent performance scaling.  Average lane utilization vs number of lanes is shown for a variety of applications, and machine scales are shown in Figure \ref{fig:lane-utilization}.  TC shows high utilization through 256 nodes.  Both PageRank and BFS show high but declining utilization -- they are system-network injection-rate limited (4.4 TB/s/node).  %For comparison, we measured single GPU/CuGraph TC core utilization at 3\%.
UpDown's efficient fine-grain parallelism mechanisms convert this lane utilization into high performance.

% \begin{figure}[h]
%     \centering
%     % \aac{add "(nodes)" to x-axis}
%     %\includegraphics[width=0.95\linewidth]{figures/utilization_overlay_boxplot_4400bw.pdf}
%    %\includegraphics[width=0.95\linewidth]{figures/lane_utilization_4400_bw.pdf}
%     \includegraphics[width=0.75\linewidth]{figures/utilization_overlay_boxplot_4400bwv4.pdf}
%     \caption{Average Lane utilization vs Machine size (nodes).  Whisker plots show 10th and 90th percentiles.} %(\jiya{need to choose one.}}
%     \label{fig:lane-utilization}
% \end{figure}

Finally, a critical difficulty in traditional CPU and GPU architectures is tolerating memory and communication latency.  UpDown exploits the massive expressed parallelism to maintain performance, even if network latency increases.  This capability is shown in Figure \ref{fig:latency-tolerance}, which shows lane utilizations in a 128-node, 256K lane machine if the network latency is increased from its nominal 0.5 $\mu$s to 8 $\mu$s.  The same UpDown program can run efficiently in all of these cases, of course, trading parallelism to hide the latency.

\begin{figure*}[t]
    \centering
    \begin{minipage}{0.49\textwidth}
        \centering
        \includegraphics[width=\linewidth]{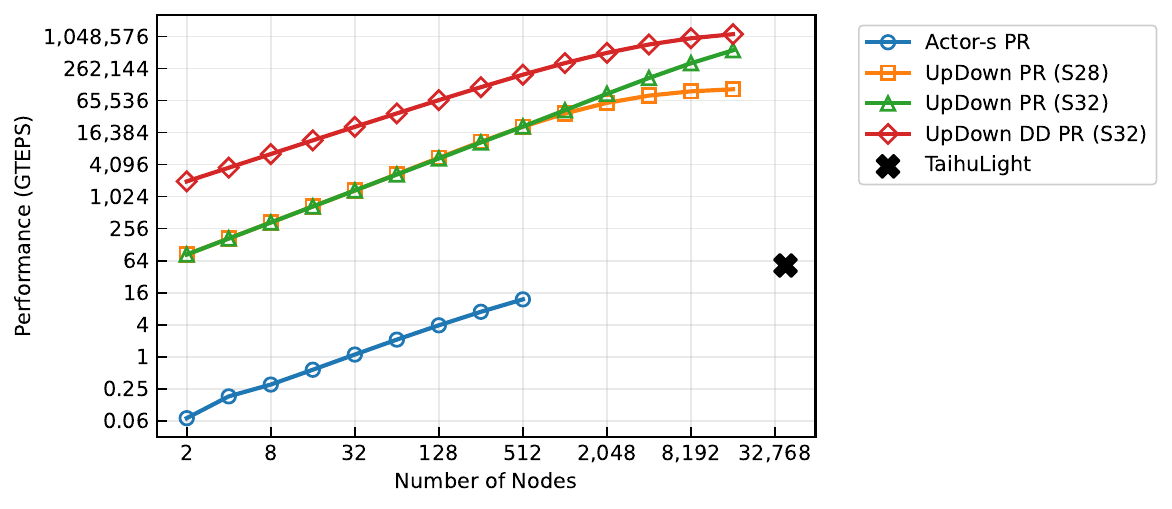}
        \caption{GTEPS for PR and {\it effective} GTEPS Data-Driven PR for \SysName (2-16K nodes). Comparison to Perlmutter and TaihuLight.
Scale 28 and 32 ER graphs.}
        \label{fig:pr-isopower-speedup}
    \end{minipage}
    \hfill
    \begin{minipage}{0.49\textwidth}
        \centering
        \includegraphics[width=\linewidth]{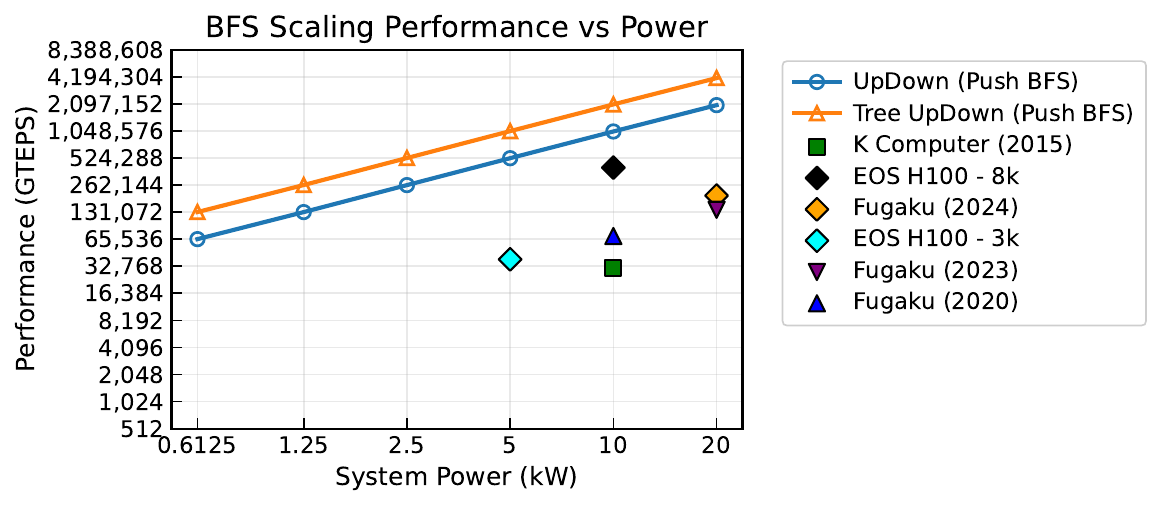}
        \caption{BFS Speedup: \pname (Scale 40) compared to notable Graph 500 results, see the discussion of Tree UpDown in Section~\ref{sec:discussion}.}
        \label{fig:bfs-isopower-speedup}
    \end{minipage}

\end{figure*}

\section{Discussion and Related Work} % Results (Other systems)}
\label{sec:other}
\label{sec:discussion}
\label{sec:related}

The claim for our paper is that UpDown enables fine-grained graph algorithms to be fast and scalable. The algorithms implemented for UpDown are work-efficient algorithms. For instance, the BFS and PageRank computations only look at each edge once. The k-core and connectivity implementations also only visit each edge once.  Moreover, we can take advantage of work-reducing algorithms like data-driven PageRank to drive effective performance higher. The algorithms that we have implemented on UpDown are fairly standard and far from exotic. \emph{We want to stress, again, that it is the same implementation that has high performance on one node and 16,384 nodes.}  We also believe that these algorithms may be made even faster by future optimization. The UpDown architecture itself demonstrates that there is a possible world of fine-grained algorithms that are both fast and scalable. 

The observation that fine-grained graph algorithms can be fast on appropriate architectures is well established (e.g., \ cite {Bader-mta2,Catalyrek2012,Subramanian2017}). As mentioned in Section II, however, accelerators will only scale as well as the interconnection. They are also often difficult to program. 

Indeed, these are the two key challenges for scale-out graph systems: programming and networking. Scalable computing and database systems, including Pregel, Giraph, and PowerGraph, indeed scale to large data problems~\cite{Malewicz-2010-Pregel,Gonzalez-2012-powergraph,GiraphPaper,Kumar2023gpu_cpu_fusion}. However, these systems typically exhibit poorer efficiency than shared-memory graph frameworks such as Ligra and Galois~\cite{Shun-2013-ligra,Nguyen-2013-galois,McSherry-2015-cost,dhulipala2021theoretically}. 
Distributed frameworks tend to make distributed systems easier to program, at the expense of absolute performance. On the network side, the types of networks installed in traditional large-scale supercomputers are not well-suited to the communication requirements of fine-grained graph algorithms. Consequently, high performance implementations have resorted to complex algorithms. 

One system with obvious similarities to UpDown is the Cray MTA2/XMT.  The MTA2 featured fine-grained parallelism and low-overhead synchronization with hardware full-empty bits~\cite{Bader-mta2,Catalyrek2012}.  Key differences include UpDown's ability to execute a thread continuously at one instruction per clock -- good single-thread performance (vs MTA2's barrel-threading) and a scratchpad which MTA2 register pressure problems.  Finally, UpDown has ample memory bandwidth (8 bytes per 3 instructions).  These properties make UpDown much simpler to exploit for high performance.

%The MTA2 This system used barrel threaded, with synchronization through full and empty bits. UpDown, instead, gives threads exclusive access to lanes until they yield access, enabling software-defined synchronization and atomic operations. This architecture difference enables simpler algorithms. 

We now compare the scalability of UpDown on PageRank and BFS to the fastest algorithms we can find.

\paragraph{PageRank}
\pname outperforms traditional and specialized supercomputers.  
%Our results here show that \pname can scale well.
%We compare our PageRank algorithms to results on the 
In Figure \ref{fig:pr-isopower-speedup}, we show \pname outperforms the Perlmutter supercomputer \cite{Elmougy2023}, using an \ErdosRenyi graph by 2000x.  The best reported numbers are labeled Actor-s.  
%, has dramatically lower performance per node.  %scales well to 64 nodes, then tapers off. 
%Because the node power for \pname and Perlmutter are similar, 
The per-node comparison graph (Figure~\ref{fig:pr-isopower-speedup}) is roughly equivalent to an ISO-power comparison.  
Shentu on Taihulight was run with multi-petabyte graphs
%We also compare to ShenTu performance studies\
\cite{Lin2018}.  TaihuLight is a general-purpose, CPU-based, extreme-scale system (10.6M cores).  
%Compared to Perlmutter, the \pname system has a 2,000-fold performance advantage for small graphs -- for 
\pname outperforms TaihuLight (52 GTEPS for scale 34 graph) by 17,000-fold.  \pnamenospace's advantage on small graphs increases for larger systems due to \pnamenospace's superior scaling record performance at 684K  GTEPS at 10MW (full scale for the \pname system).
TaihuLight achieves 1,984 GTEPS on a scale 40 Kronecker graph with a 15MW system.  Here, \pname is 345x faster and 500-fold superior (ISO-power). 
Data-driven PageRank achieves a higher effective GTEPS of 9.6M, 14x faster, showcasing the power of programmability.

\paragraph{BFS and Graph500 winners}
The Graph500 benchmark provides a common benchmark for comparing implementations and performance of BFS. Software optimization is known to play a key role in this benchmark, dating back to initial realizations of how much better push-pull BFS is than standard BFS~\cite{Beamer2012}. Another recent innovation in software size is the forest-of-trees approach, or 2-core optimization, discovered by the Fugaku team~\cite{Arai2024}. This, combined with other simple software tweaks, resulted in a 2x performance increase for the Fugaku result from 2023 to 2024. Our UpDown software is not nearly as mature as these implementations and currently uses only a simple \emph{push}- based algorithm. Consequently, we project two lines for UpDown's performance: the current projected results and a software-optimized version that includes this forest-of-trees optimization. 

The \pname system has both high efficiency and excellent scalability, reaching 988K  GTEPS (10MW, full \pnamenospace) and 1.94M GTEPS (20MW). The hypothetical software optimizations double this to 2M GTEPS at 10MW and 4M GTEPS at 20MW.  Comparison to 2020-25 Graph 500 winners in Figure~\ref{fig:bfs-isopower-speedup} show 5x vs Fugaku  \cite{Arai2024} (\#1 on Graph 500 in May 2025), 10x ISO-power.  Graph 500 November 2025's winner is a cluster of Nvidia H100 superpods (1024-nodes x 8 GPUs each) that reached a 410k GTEPS at a power of around 10MW. UpDown achieves 2.3x this performance at similar power, which algorithmic improvements (e.g., push-pull, as in Figure~\ref{fig:bfs-perf}) will improve this to over 5-8x. 

\section{Summary and Future Work}
\label{sec:summary}

% \ivy{@Ivy fix the summary claims and future work}
We studied a co-designed system for graph processing that exploits the full fine-grained parallelism expressible in graph applications. 
%We described the \SysName architecture 
% that consists of over 30 million highly efficient cores in 16,384 nodes coupled with a high bandwidth network interconnect. We 
%and conducted 
Detailed simulations show 28x (geomean) single-node speedup compared to H200 GPU.  Detailed simulation and projection studies show excellent speedups to 256 nodes and projections for 16K nodes, indicating that the co-designed architecture can exceed the performance of the fastest existing systems by up to 100-fold on these applications.  
%The codesigned architectures can achieve dramatically more performance on these applications.
%of the system in the simulator up to 256 nodes (524,288 cores) and evaluated the performance on a mixture of generated and real-world graphs. This showed that the system is remarkably efficient at utilizing the fine-grained parallelism in the graph application, \ivy{with a speedup of 178x and 377x self-normalized to one node performance and achieves a maximum GTEPS of 33,843 and 18,231 for PageRank and BFS, respectively.} We took the simulated performance of the algorithms and then created a projection model to forecast the system performance out to the full 16,384 nodes and 33 million system cores. This showed that the system achieved results that exceeded 1 million GTEPS for both BFS and PageRank. We compared this performance level to recently proposed systems on the Graph500 list for BFS and other descriptions of highly scalable PageRank algorithms, which shows \SysName is about 15x (BFS) and 2000x (PageRank) the performance of existing systems. \ivy{@Charlie and @Andrew can you double check if this is correct? it seems to take up a lot of space. Should we switch to a more high-level description?}
% 

% Future Work 
Interesting directions for future work include
%:  (1) a deeper study of application needs and system network performance because of the irregularity induced by real-world graphs, . 
%utilization, scrutinizing for potential bottlenecks.
%Second, full studies on a detailed design (nearly complete as part of the \AGILE) to study this architecture in more depth.    
%AGILE program \cite{AGILE}.  
%Finally,
(i) a deeper study of 
%we have only scratched the surface of what is possible with a 
the natural programming model (directly vertex and edge parallelism) and    
(ii) explorations of even more challenging graph computing problems and sophisticated algorithms.
%, and (iii) comparison against other graph database uses.

%more problems and more sophisticated algorithms

\excise{
\begin{section}
This research is supported by the Office of the Director of National Intelligence (ODNI), Intelligence Advanced Research Projects Activity (IARPA), through the Advanced Graphical Intelligence Logical Computing Environment (AGILE) research program, under Army Research Office (ARO) contract number W911NF22C0082. The views and conclusions contained herein are those of the authors and should not be interpreted as necessarily representing the official policies or endorsements, either expressed or implied, of the ODNI, IARPA, or the U.S. Government.

This work is also supported in part by NSF Grant CNS-1907863 and a Computation Innovation Fellows Award.  Thanks also to the entire UChicago UpDown team, and to our collaborators at Purdue University and Tactical Computing Laboratories.
\end{section}
}

% \newpage
\bibliographystyle{IEEEtran}
\bibliography{refs,graph-accels}

% \vspace{12pt}
% \color{red}

% \input{to_fix}

% \input{background}

\end{document}